\documentstyle[a4,12pt,epsf]{article}
\epsfverbosetrue
\newcommand{\plus}{\makebox[15pt][c]{$+$}}
\newcommand{\minus}{\makebox[15pt][c]{$-$}}
\newcommand{\figurebox}[2]{\fbox{\vbox to#2in{\hbox to #1in{\hfil} \vfil}}}
\newcommand{\errr}[2]{\raisebox{0.08em}{\scriptsize {$\;\begin{array}{@{}l@{}}
                          \plus\makebox[0.9em][r]{#1} \\[-0.12em]
                          \minus\makebox[0.9em][r]{#2}
                        \end{array}$}}}
\newcommand{\err}[2]{\raisebox{0.08em}{\scriptsize {$\;\begin{array}{@{}l@{}}
                          \plus\makebox[0.55em][r]{#1} \\[-0.12em]
                          \minus\makebox[0.55em][r]{#2}
                        \end{array}$}}}
\newcommand{\er}[2]{\raisebox{0.08em}{\scriptsize {$\;\begin{array}{@{}l@{}}
                          \plus\makebox[0.15em][r]{#1} \\[-0.12em]
                          \minus\makebox[0.15em][r]{#2}
                        \end{array}$}}}
\newcommand{\eqn}[1]{Equation~(\ref{#1})}

\newcommand{\fig}[1]{Figure~\ref{#1}}
\newcommand{\tab}[1]{Table~\ref{#1}}
\newcommand{\bm}[1]{\mbox{\boldmath ${#1}$}}

\newcommand{\beq}{\begin{equation}}
\newcommand{\eeq}{\end{equation}}

\newcommand{\kcrit}{\mbox{$\kappa_{\rm crit}$}}
\newcommand{\kstrange}{\mbox{$\kappa_s$}}

\newcommand{\gev}{{\rm GeV}}
\newcommand{\msbar}{\overline{{\rm MS}}}

\begin{document}

\begin{titlepage}

\begin{flushright}
Edinburgh Preprint: 93/524\\
Southampton Preprint SHEP 92/93-16\\
May 27, 1993
\end{flushright}

\vspace*{5mm}

\begin{center}
{\Huge The Light Hadron Spectrum and Decay Constants in Quenched
Lattice QCD}\\[15mm]
{\large\it UKQCD Collaboration}\\[3mm]

{\bf C.R.~Allton~\footnote{Present address: Dipartimento di
Fisica, Universit\`a di Roma {\em La Sapienza}, 00185 Roma,
Italy.}, L.~Lellouch, C.T.~Sachrajda, H.~Wittig}\\
Physics Department, The University, Southampton SO9~5NH, UK

{\bf R.M.~Baxter, S.P.~Booth, K.C.~Bowler, D.S.~Henty, R.D.~Kenway,
C.~McNeile\footnote{Present address: Dept.~of
Physics \& Astronomy, University of Kentucky, Lexington KY 40506,
USA}, B.J.~Pendleton, D.G.~Richards, J.N.~Simone, A.D.~Simpson}\\
Department of Physics, The University of Edinburgh, Edinburgh EH9~3JZ,
Scotland

\end{center}
\vspace{5mm}
\begin{abstract}

We present results for light hadrons composed of both degenerate
and non-degenerate quarks in quenched lattice QCD.  We calculate
masses and decay constants using 60 gauge configurations with an
$O(a)$--improved fermion action at $\beta = 6.2$.  Using the
$\rho$ mass to set the scale, we find hadron masses within two to
three standard deviations of the experimental values (given in
parentheses): $m_{K^*}=868\er{9}{8}$~MeV (892~MeV),
$m_{\phi}=970\err{20}{10}$~MeV (1020~MeV),
$m_N=820\err{90}{60}$~MeV (938~MeV),
$m_\Delta=1300\errr{100}{100}$~MeV (1232~MeV) and
$m_\Omega=1650\err{70}{50}$~MeV (1672~MeV).  Direct comparison
with experiment for decay constants is obscured by uncertainty in
current renormalisations.  However, for ratios of decay constants
we obtain $f_K/f_\pi=1.20\er{3}{2}$ (1.22) and
$f_\phi/f_\rho=1.13\er{2}{3}$ (1.22).

\end{abstract}

\end{titlepage}


\section{Introduction}
Within the quenched approximation, it is currently possible to
study lattice QCD numerically in a box of linear size around 2~fm,
with a lattice spacing of less than 0.1~fm, corresponding to a
cutoff above 2~GeV.  Although extrapolation to the chiral limit is
still a necessary ingredient when $u$ and $d$ quarks are involved,
such lattices allow the direct simulation of hadrons containing
$s$ quarks, at the cost of fixing one additional mass parameter.
A significantly wider range of physical quantities thereby becomes
calculable, with which to probe the reliability of lattice QCD.

In this paper, we extend an earlier
study~\cite{UKQCD_hadrons_lett,UKQCD_hadrons_npb} of light hadron
masses and decay constants to include the effects of
$SU(3)$-flavour-symmetry breaking.  The earlier study was based on
18 configurations, and the correlation functions were evaluated
for hadrons composed of degenerate quarks, for five different
values of the quark mass, using both the Wilson and the clover
fermion actions.  The results presented in this paper were
obtained from our complete data set of 60 configurations, using
the clover fermion action, for three of the five previously-used
quark masses.  In order to study flavour-symmetry-breaking
effects, we construct mesons using all possible quark-mass
combinations.

In the Wilson formulation, the bare quark mass, $m$, is given in
terms of the hopping parameter $\kappa$ by
\begin{equation}
m = \frac{1}{2}\left(\frac{1}{\kappa} - \frac{1}{\kcrit}\right).
\end{equation}
\kcrit\ is the value of the hopping parameter at zero quark mass,
which is taken to be the point at which the mass of the
pseudoscalar meson, $m_{P}$, vanishes and the quark and antiquark
are degenerate.  With each quark flavour, we need to associate a
value of $\kappa$ corresponding to its experimentally observed
mass.  It is a good approximation to take the physical light quark
($u$ and $d$) masses to be zero, i.e.\ to set $\kappa_u = \kappa_d
= \kcrit$.  Here we are interested in computing the effect of
$SU(3)$-flavour-symmetry breaking on the spectrum and decay
constants, and so we need to associate a non-zero mass with the
strange quark.  The corresponding $\kappa_s$ can be determined,
for example, by first extrapolating the vector meson mass,
$m_V(\kappa_1,\kappa_2)$, to $\kappa_1=\kappa_2=\kcrit$ and then
fitting the data for the ratio
$m^2_{P}(\kappa_1,\kappa_2)/m^2_V(\kcrit,\kcrit)$ to some function
of the two quark masses, extrapolating in $\kappa_1$ to \kcrit\
and using $\kappa_2$ to fix the ratio to the experimental value of
$m_K^2/m_\rho^2$.  Alternatively, $\kappa_s$ may be determined
from the degenerate-quark data alone, avoiding the second chiral
extrapolation, by using the ratio $m_{\Omega}/m_{\rho}$.

This procedure requires some assumption about how the hadron
masses depend on the quark masses.  In most previous calculations
(see, for
example,~\cite{UKQCD_hadrons_npb,fucito,lipps,MM_86,weingarten_spectrum}),
it has been assumed that the pseudoscalar meson mass obeys the
PCAC relation
\begin{equation}
m_{P}^2(\kappa_1,\kappa_2) = b_{P}\left( \frac{1}{2\kappa_1}
				+\frac{1}{2\kappa_2}
				-\frac{1}{\kcrit}\right),
\label{eq:dg_two_kappas}
\end{equation}
and that the vector meson mass obeys
\begin{equation}
m_V(\kappa_1,\kappa_2)  =  a_V + b_V\left( \frac{1}{2\kappa_1}
                                +\frac{1}{2\kappa_2}
                                -\frac{1}{\kcrit}\right).
\label{eq:dgV_two_kappas}
\end{equation}
The original work of Martinelli et al.~\cite{omero} supported this
assumption, albeit on the basis of rather limited statistics, by
combining quark propagators computed for different values of
$\kappa$.  More recent studies of strange
hadrons~\cite{iwasaki,loft_degrand} have similarly utilised
non-degenerate quarks, and this is the procedure that we adopt
here.

\begin{figure}[htbp]
\begin{center}
\leavevmode
\epsfysize=250pt
  \epsfbox{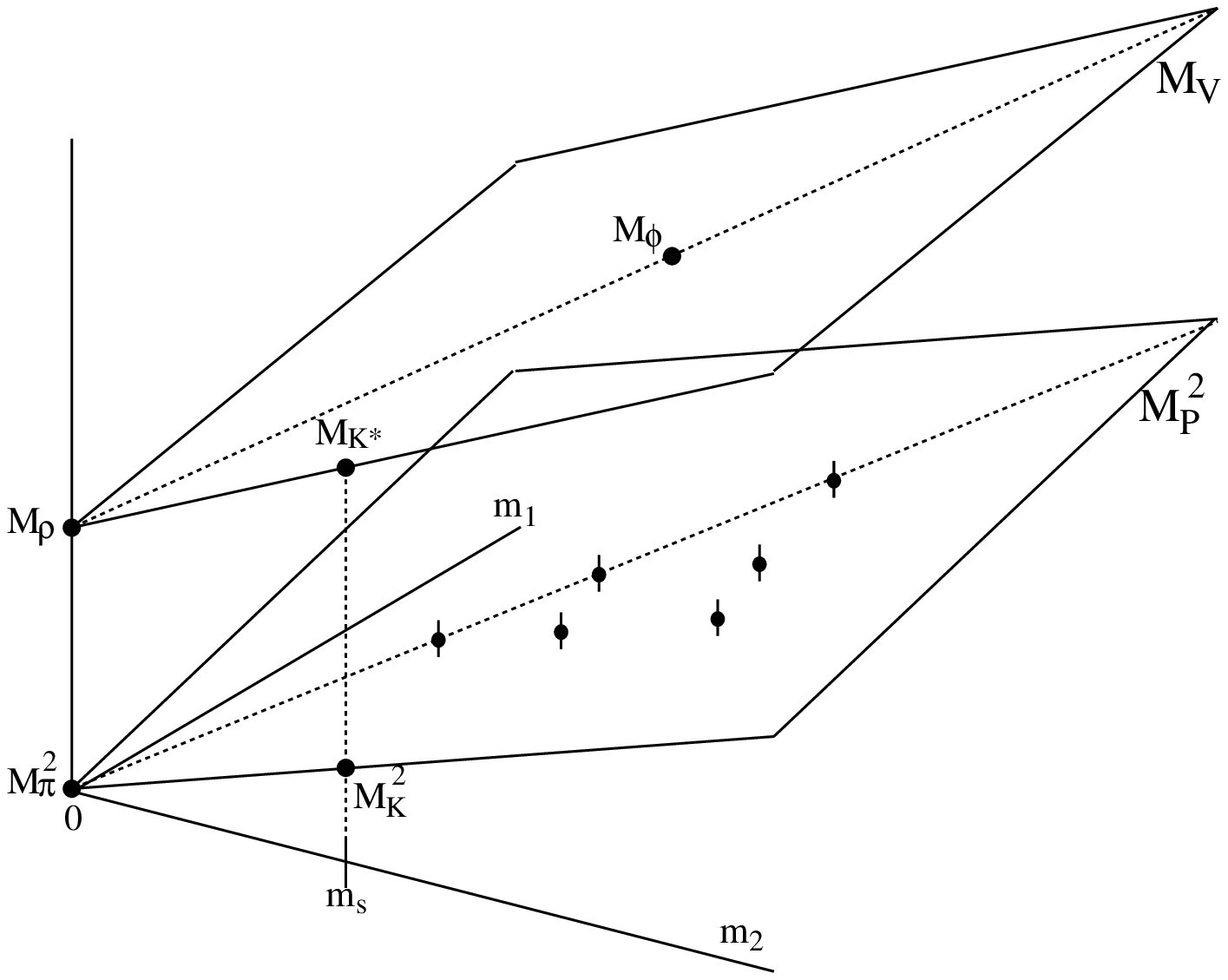}
\end{center}
\caption{Schematic of the fit planes defined by
Equation~(\protect\ref{eq:ndg_fit_form}).}
\label{drawing}
\end{figure}

We fit our data for $m^2_{P}$, $m_V$, $f_{P}$, etc., to the
following function of the two quark masses, $m_1$ and $m_2$,
\begin{equation}
a_1 + \frac{a_2}{2} \; (m_2 + m_1) + \frac{a_3}{2} \; |m_2 - m_1|
\label{eq:ndg_fit_form}
\end{equation}
to test the assumption that $a_3 = 0$ in
Equations~(\ref{eq:dg_two_kappas}) and (\ref{eq:dgV_two_kappas}).
The expression~(\ref{eq:ndg_fit_form}), or its equivalent with
$a_3 = 0$, defines a plane through the data, as shown in Figure
{}~\ref{drawing}.  By means of a combination of extrapolation and
interpolation using such fits, we are able to calculate the masses
and decay constants of the $K$, $K^*$ and $\phi$ mesons and the
mass of the $\Omega$ baryon, in addition to those of the usual
light hadrons.

In Section 2, we summarise our statistics and present details of
our fitting criteria and procedures.  Our main results are given
in Section 3, where we describe the chiral extrapolations, the
interpolation to the strange quark mass, and the masses and decay
constants in physical units. Section 4 contains our conclusions.



\section{Computational Details}
%
\subsection{Statistics}

We have analysed 60 gauge field configurations at an inverse coupling
$\beta = 6.2$ on a lattice of size $24^3 \times 48$.  The gauge
configurations and quark propagators were produced on the 64-node
i860 Meiko Computing Surface at the University of Edinburgh.  The
$SU(3)$ gauge fields were generated using the Hybrid Over-Relaxed
algorithm, defined in reference~\cite{UKQCD_hadrons_npb}.  The
gauge configurations are separated by 2400 sweeps, beginning at
configuration 16800.  The quark propagators were calculated using
an $O(a)$-improved clover action~\cite{SW_85,HSMPR_91}
\begin{equation}
S_F^C  = S_F^W - i\frac{\kappa}{2}\sum_{x,\mu,\nu}\bar{q}(x)
         F_{\mu\nu}(x)\sigma_{\mu\nu}q(x).
\end{equation}
$S_F^W$ is the standard Wilson lattice action,
\begin{equation}
S_F^W  =  \sum _x \Biggl\{\bar{q}(x)q(x)
             -\kappa\sum _\mu\Bigl[
            \bar{q}(x)(1-\gamma _\mu )U_\mu (x) q(x+\hat\mu ) +
            \bar{q}(x+\hat\mu )(1 + \gamma _\mu )U^\dagger _\mu(x)
            q(x)\Bigr]\Biggr\}
\end{equation}
and $F_{\mu\nu}$ is a lattice definition of the field strength
tensor.  We have computed propagators at three values of $\kappa$,
0.14144, 0.14226 and 0.14262, using an over-relaxed minimal
residual algorithm with red-black preconditioning and point
sources and sinks.  Although there are advantages to using smeared
sources and/or sinks to extract ground-state properties from
2-point functions~\cite{UKQCD_smearing}, constraints imposed by
other parts of the UKQCD programme did not allow us this option.

We construct correlators for mesons composed of quarks of flavours
1 and 2 using the following local interpolating fields:
\begin{eqnarray}
P     &=& \bar{q}_1\gamma_5 q_2\\
A_4   &=& \bar{q}_1\gamma_4\gamma_5 q_2\\
V_i   &=& \bar{q}_1\gamma_i q_2,
\end{eqnarray}
and correlators for baryons composed of degenerate quarks using
\begin{eqnarray}
N     &=& \epsilon_{abc}(u^aC\gamma_5 d^b)u^c\\
\Delta_\mu &=& \epsilon_{abc}(u^aC\gamma_\mu u^b)u^c.
\end{eqnarray}
For the vector meson, we average our correlators over the three
polarisation states, for the nucleon we average the 11 and 22
spinor indices of the correlator, and for the $\Delta$ we project
out the spin-$\frac{3}{2}$ component and average over the four spin
projections.  Our quark propagators incorporate the rotations
required to ensure that using these interpolating fields yields
$O(a)$-improved correlators.  This and our computational procedure
are described in detail in reference~\cite{UKQCD_hadrons_npb}.

Except where explicitly stated otherwise, the errors quoted in
this paper are purely statistical and are calculated according to
the prescription:
\begin{itemize}

\item create 1000 bootstrap samples from the original dataset of
60 configurations by randomly choosing, with replacement, 60
configurations per sample;

\item for each bootstrap sample, perform all the mass fits and
extrapolations as for the original data;

\item obtain the errors on a given quantity from the $68\%$
confidence limits of the corresponding bootstrap distribution.

\end{itemize}


\subsection{Fitting Procedure}
We construct 2-point meson correlation functions from quark
propagators with all combinations of the three $\kappa$ values.
We perform least-$\chi^2$ fits to the zero-momentum,
time-symmetrised time-slice correlators to single cosh functions.
For the pseudoscalar channel, we fit over the time range $t=14$ to
$22$ for all $\kappa$ combinations.  For the vector channel, we
use the fitting range $t=13$ to $23$ for all except the heaviest
degenerate-$\kappa$ case, where we use $t=15$ to $23$.  For the
baryons, we construct 2-point correlation functions only for the
degenerate cases.  We fit the appropriate average of the forwards
and backwards, zero-momentum time-slice correlators to single
exponential functions, choosing the time ranges $t=16$ to $22$ for
the nucleon and $t=16$ to $21$ for the $\Delta$.  We take account
of time correlations in the least-$\chi^2$ fits.

\begin{figure}[htbp]
\begin{center}
\leavevmode
\epsfysize=235pt
  \epsfbox[20 30 620 600]{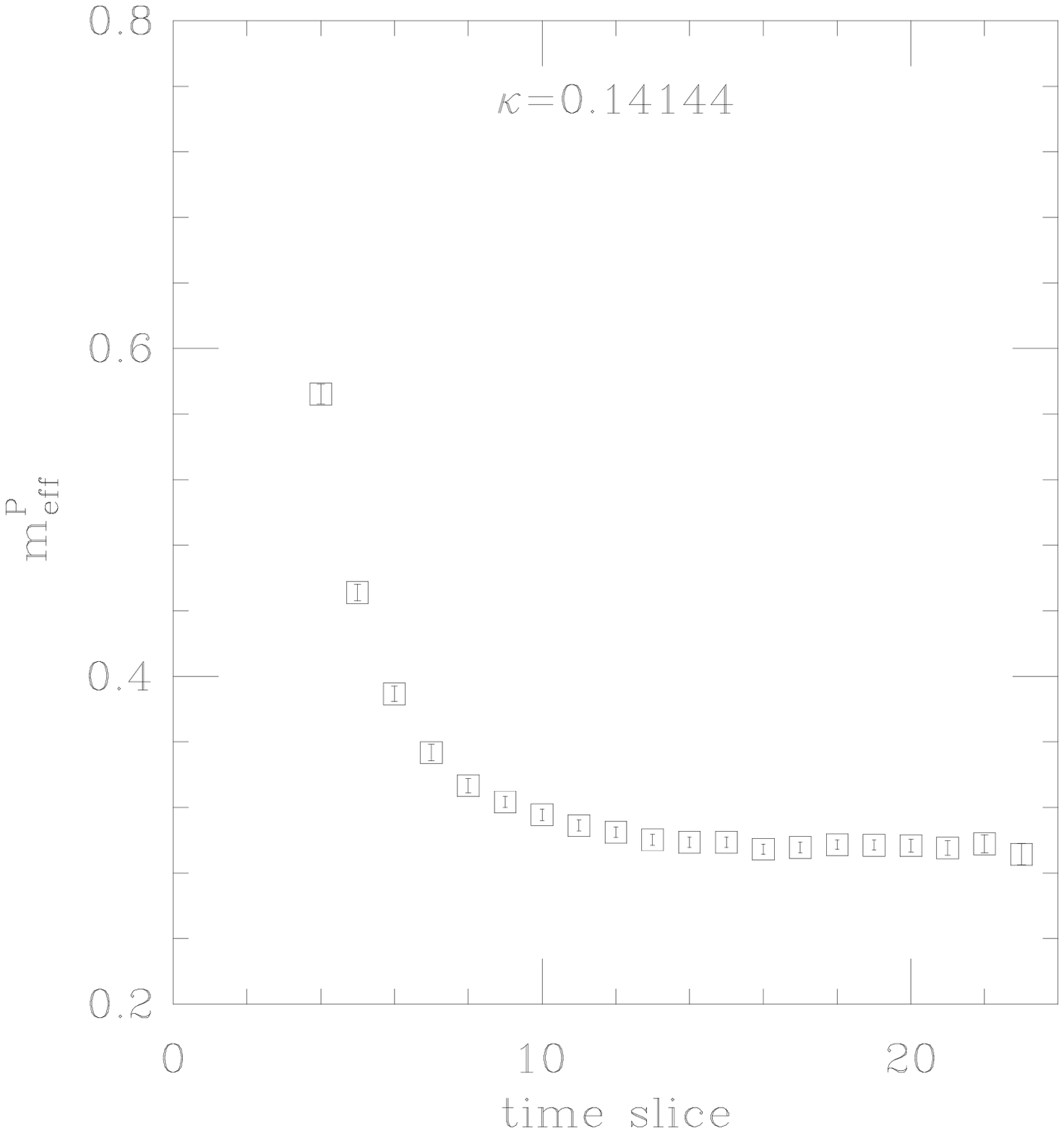}
\leavevmode
\epsfysize=235pt
  \epsfbox[20 30 620 600]{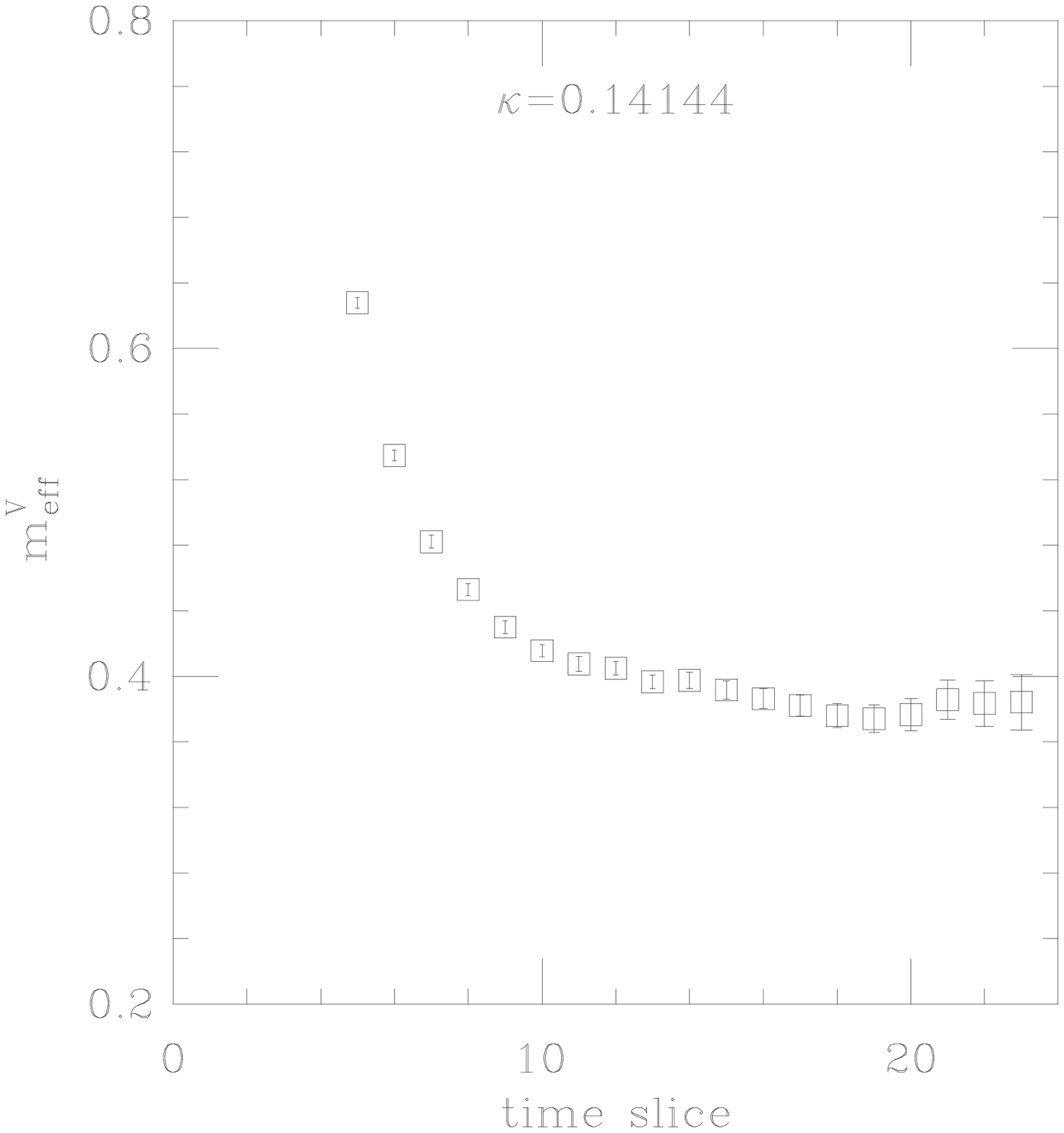}
\end{center}

\begin{center}
\leavevmode
\epsfysize=235pt
  \epsfbox[20 30 620 600]{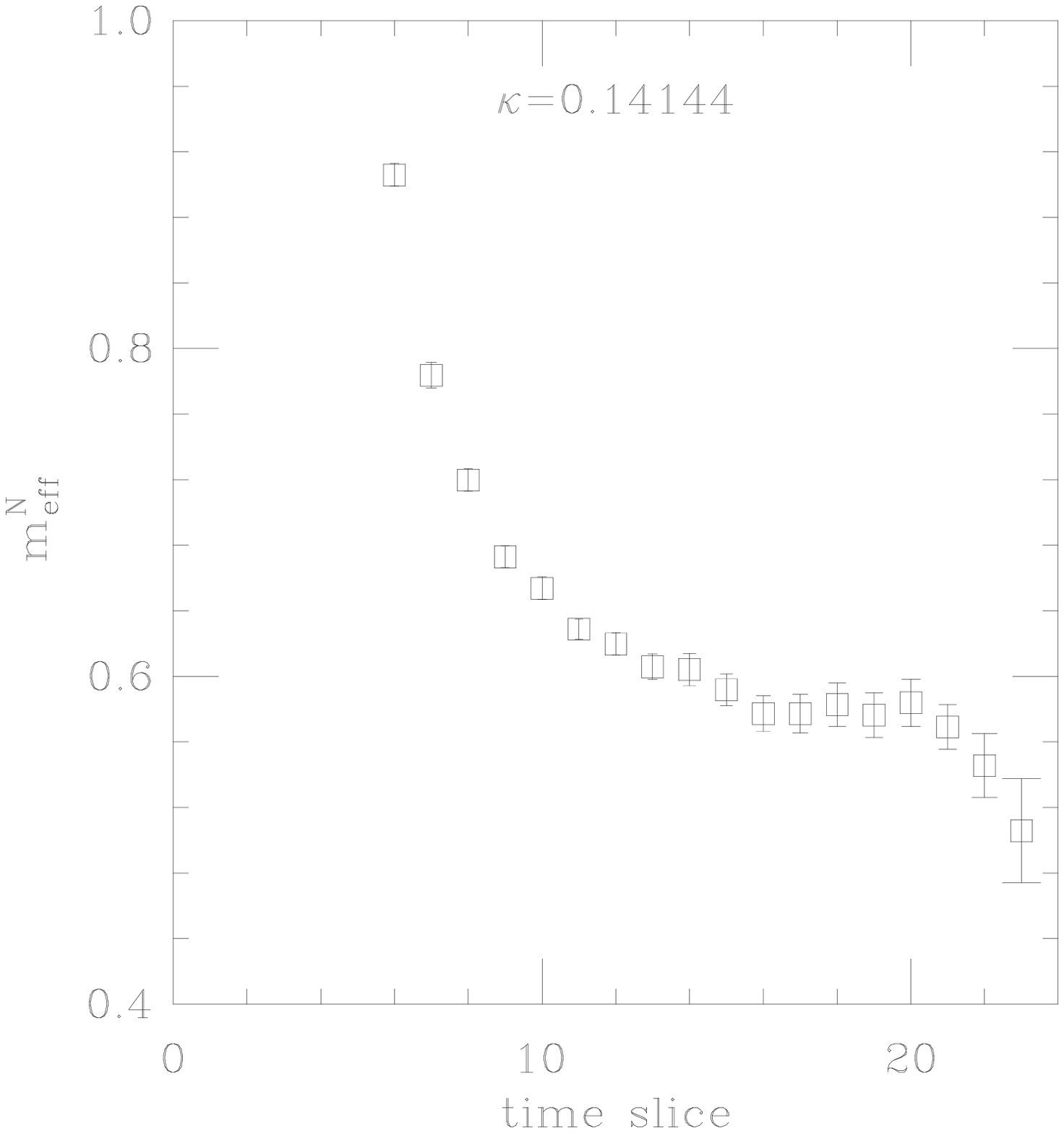}
\leavevmode
\epsfysize=235pt
  \epsfbox[20 30 620 600]{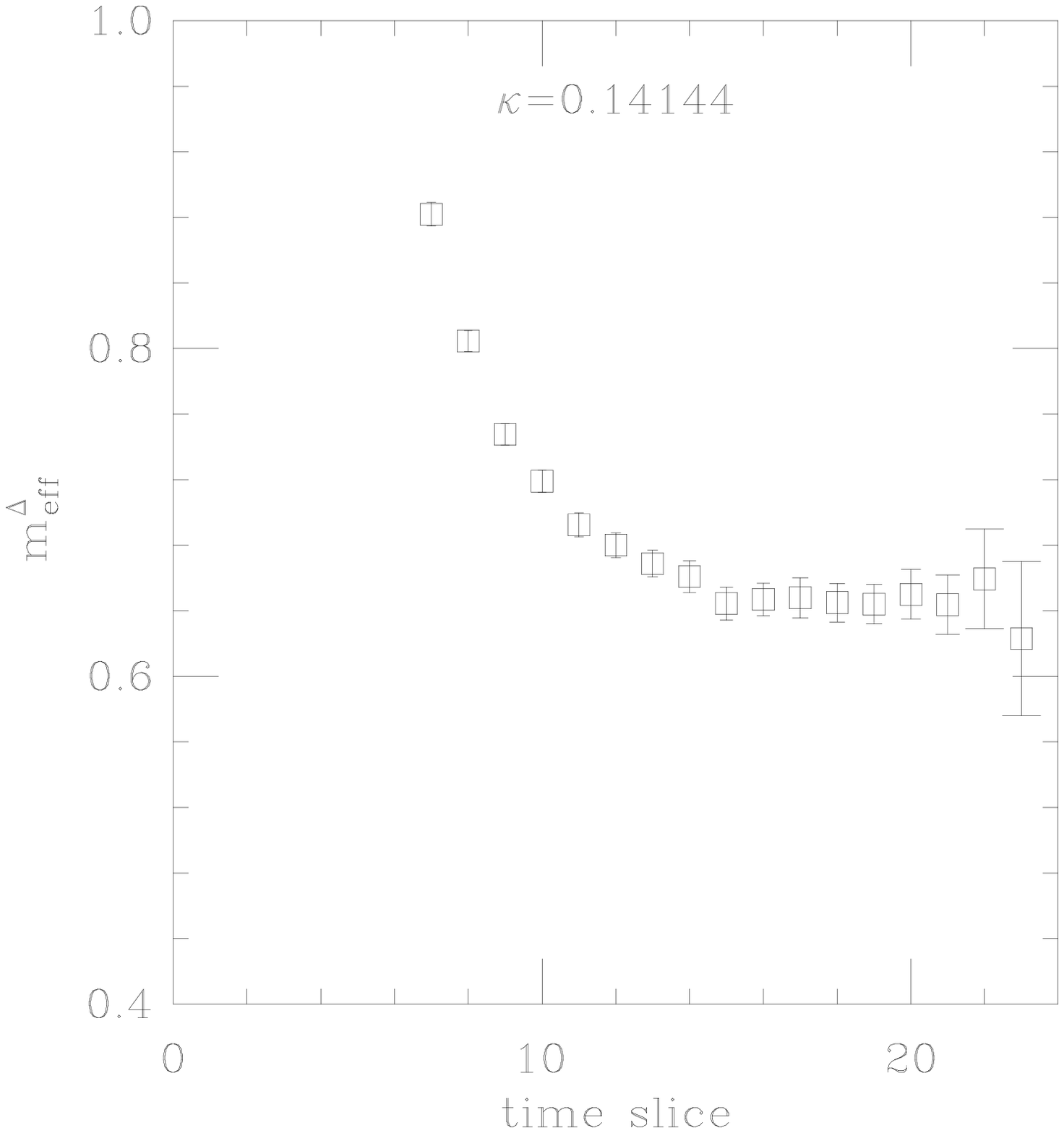}
\end{center}
\caption{Effective mass plots for the pseudoscalar, vector, nucleon
and $\Delta$ at $\kappa = 0.14144$.
}
\label{fig:eff_mass_14144}
\end{figure}

\begin{figure}[htbp]
\begin{center}
\leavevmode
\epsfysize=235pt
  \epsfbox[20 30 620 600]{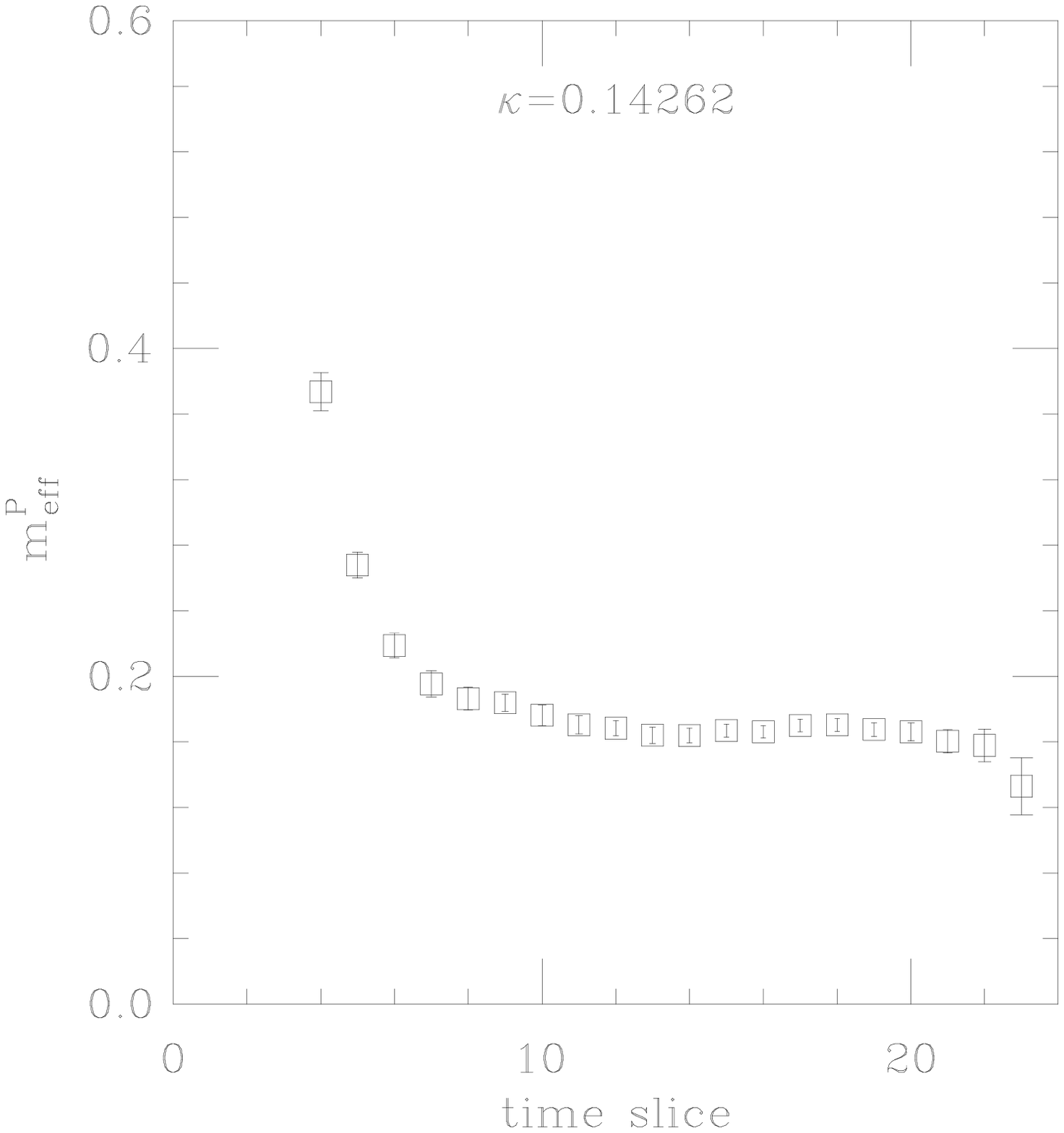}
\leavevmode
\epsfysize=235pt
  \epsfbox[20 30 620 600]{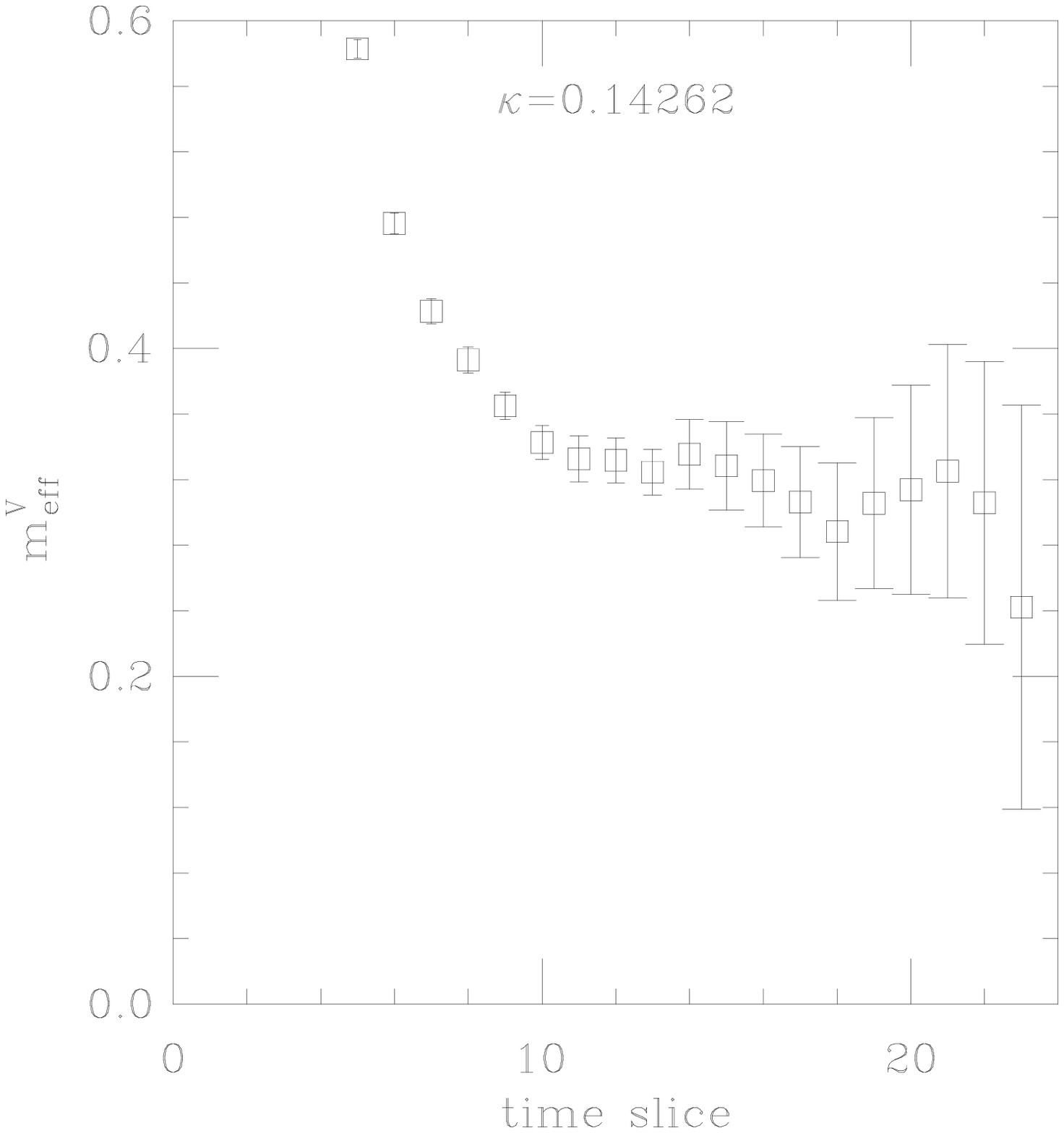}
\end{center}

\begin{center}
\leavevmode
\epsfysize=235pt
  \epsfbox[20 30 620 600]{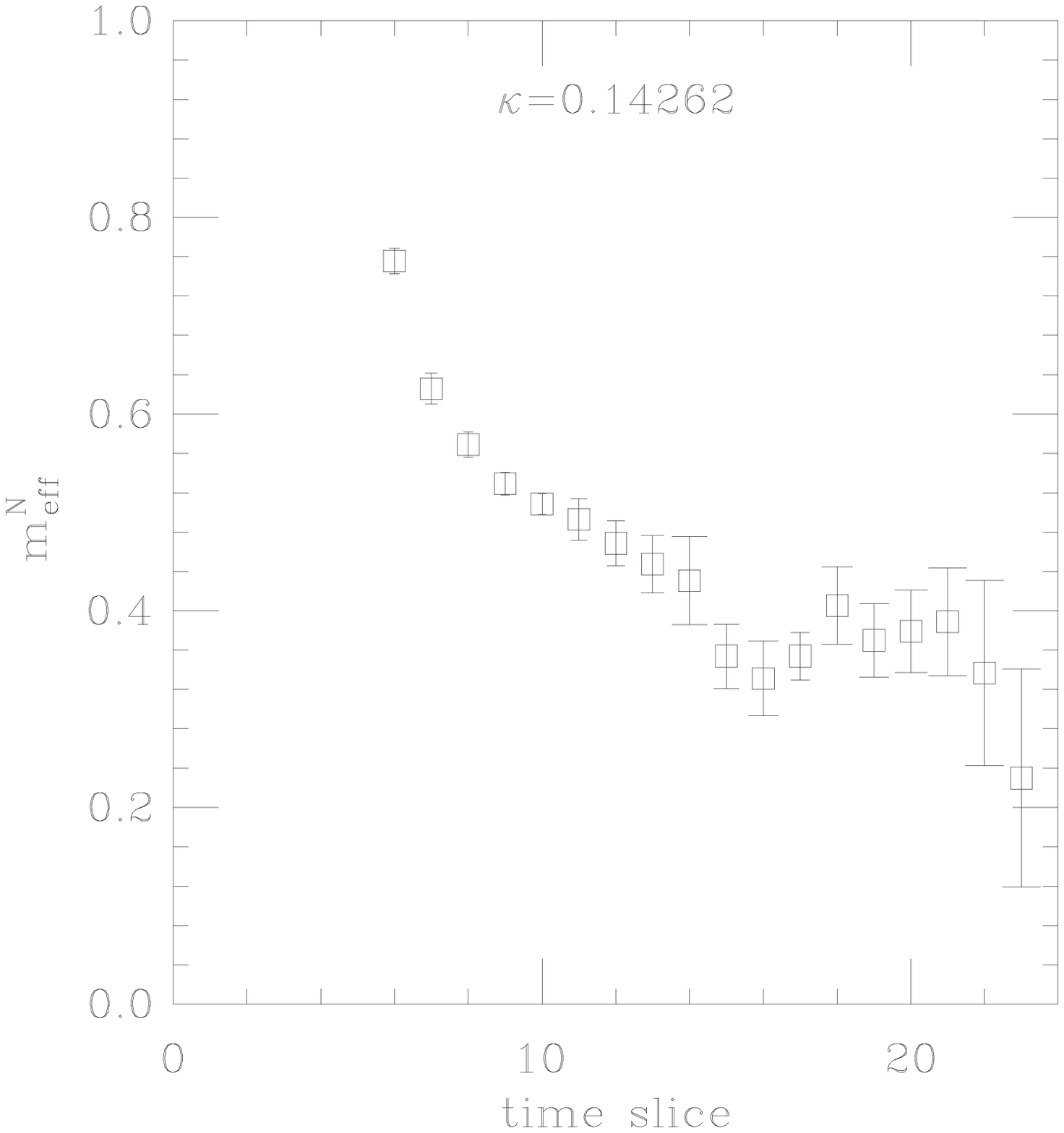}
\leavevmode
\epsfysize=235pt
  \epsfbox[20 30 620 600]{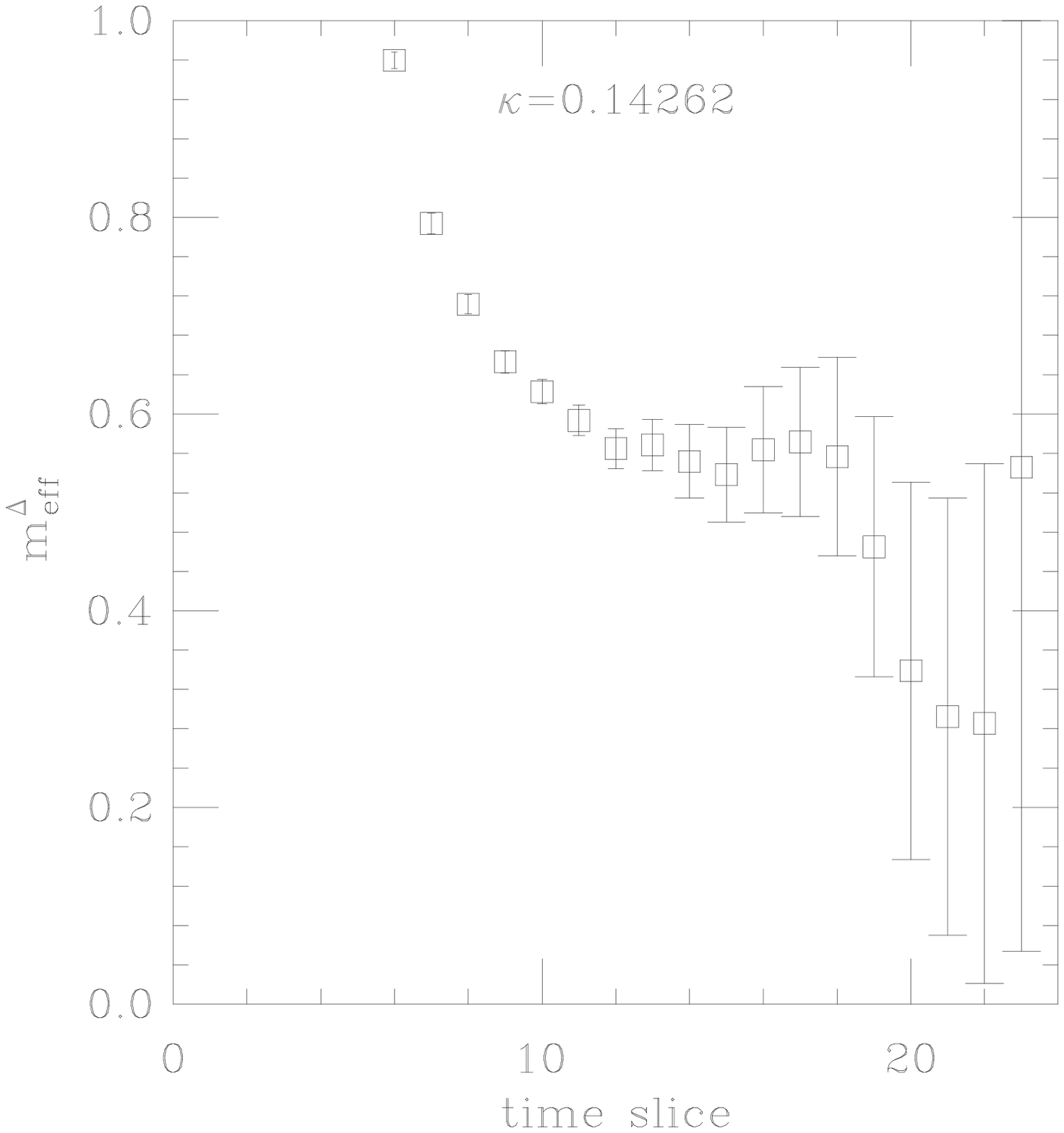}
\end{center}
\caption{Effective mass plots for the pseudoscalar, vector, nucleon
and $\Delta$ at $\kappa = 0.14262$.
}
\label{fig:eff_mass_14262}
\end{figure}

We carried out an extensive investigation of the most appropriate
fitting ranges, before arriving at the above choices.  With
reference to the effective mass plots in
Figures~\ref{fig:eff_mass_14144} and ~\ref{fig:eff_mass_14262}, we
fixed $t_{\rm max}$ to be as large as possible and reduced $t_{\rm
min}$ until the $\chi^2/$dof showed a significant increase.  In
this way, we attempted to fit as many time slices as possible.  We
followed the reasoning of reference~\cite{QCDPAX_92} in fitting
our data as far out as possible to avoid contamination from
excited states at earlier times.
 %
%
 %

\begin{figure}[htbp]
\begin{center}
\leavevmode
\epsfysize=300pt
  \epsfbox[20 30 620 600]{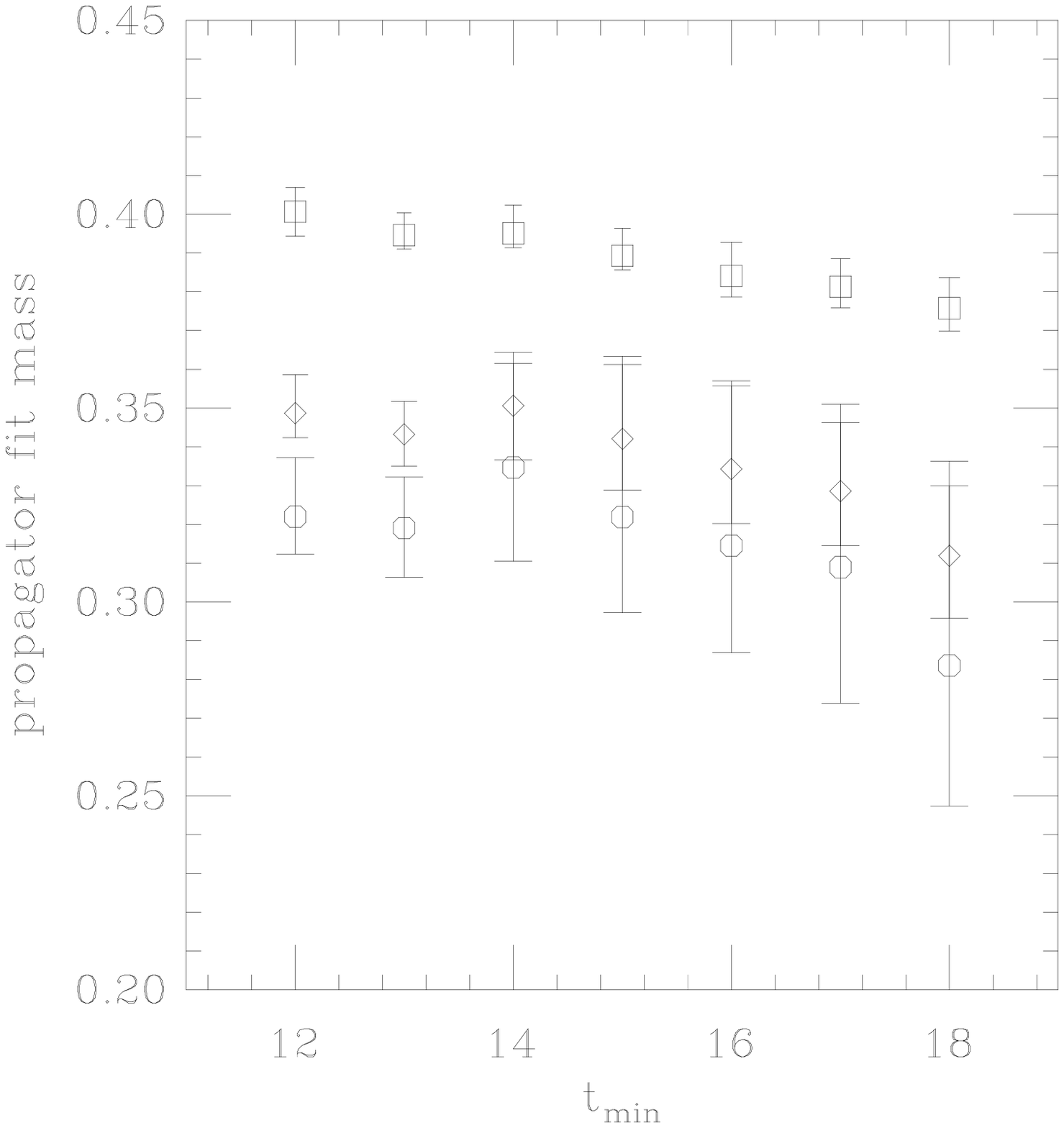}
\end{center}

\begin{center}
\leavevmode
\epsfysize=300pt
  \epsfbox[20 30 620 600]{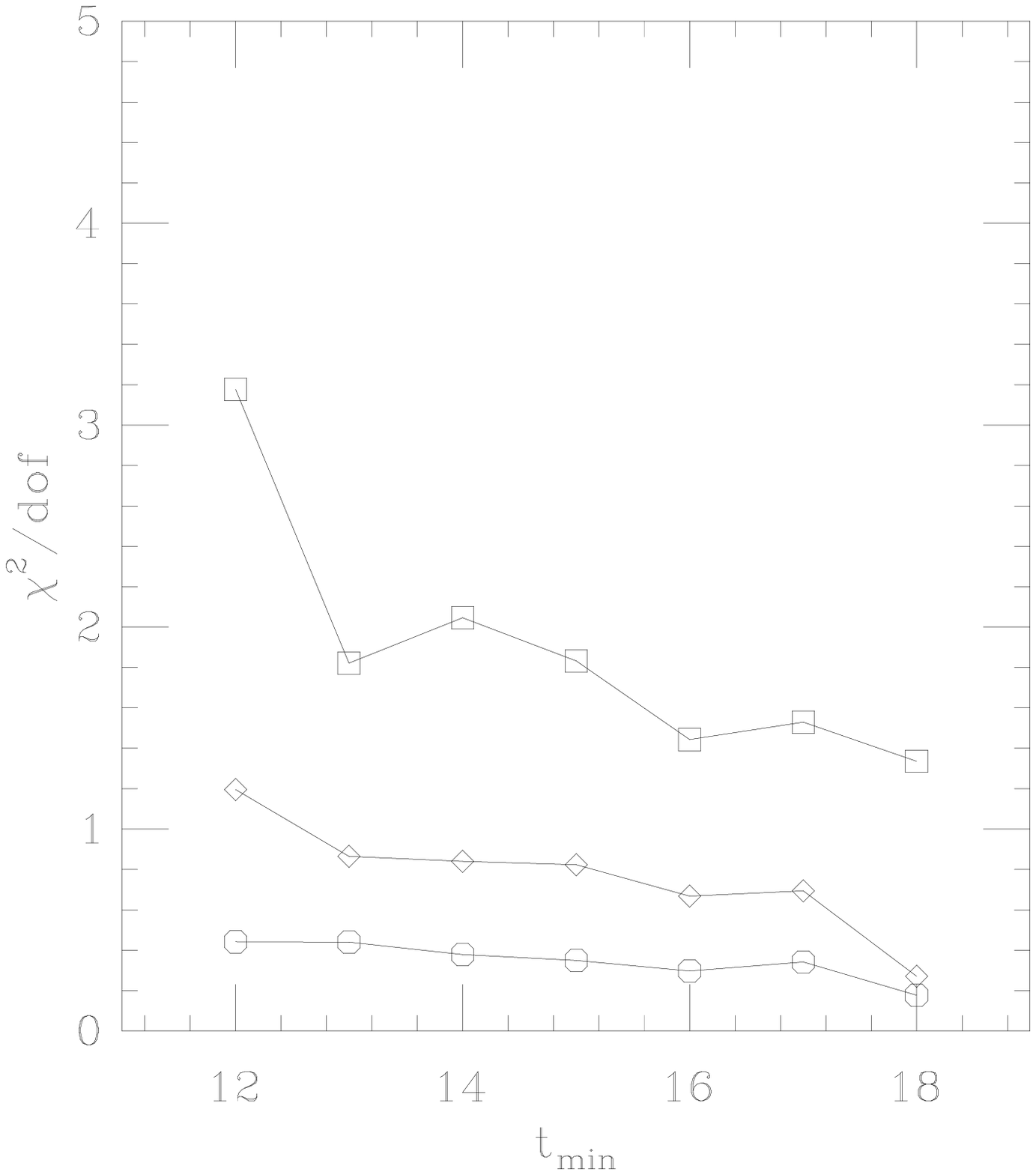}
\end{center}
\caption{Fit stability plots for the vector meson.  The top graph
shows the variation of fit mass with $t_{\rm min}$, fixing $t_{\rm
max}=23$.  The bottom graph shows the corresponding $\chi^2/$dof.
The different symbols refer to different $\kappa$ values:
squares~=~0.14144, diamonds~=~0.14226, circles~=~0.14262.}
\label{fig:V_stability}
\end{figure}

In \fig{fig:V_stability} we show the variation of the
degenerate-quark vector meson mass estimates and of the
corresponding $\chi^2/$dof with the position of the first
time slice, $t_{\rm min}$, in a variable window, fixing $t_{\rm
max} = 23$.
 %
%
%
 %
We observe a slow monotonic decrease from $t_{\rm min}=14$
onwards, at all $\kappa$ values, although there is no significant
variation in $\chi^2/$dof from $t_{\rm min}=13$ and the mass
estimates agree within statistical errors for the two lightest
$\kappa$ values.  The case of the heaviest $\kappa$ value is more
problematic given the small statistical errors.  We adopt the
compromise choice of $t_{\rm min}=15$ for this, which is
consistent with a wider range of other choices of $t_{\rm min}$.
We attempt to quantify in \tab{tab:V_stability} the systematic
error arising from the choice of fit range, by taking the lowest
and highest mass estimates from fits with values of $t_{\rm min}$
acceptable according to the above criteria.  We note that the
monotonic decrease in our mass estimates with increasing $t_{\rm
min}$ is reflected in the asymmetry of the systematic error
estimates towards lower mass values.  However, beyond this, we
feel unable to quote quantitative estimates of this error in our
results.
 %
\begin{table}
\begin{center}
\begin{tabular}{|c||c|c|c|c|c|}
\hline
 & & & & \multicolumn{2}{|c|}{error} \\
\cline{5-6}
$\kappa$ & $m_V^{\rm high}$ & $m_V^{\rm low}$ & $m_V$ & statistical &
systematic \\
\hline
0.14144  & 0.395 & 0.376 & 0.389 & $+0.007 -0.006$ & $+0.006 -0.013$ \\
0.14226  & 0.351 & 0.312 & 0.343 & $+0.009 -0.007$ & $+0.008 -0.031$ \\
0.14262  & 0.335 & 0.284 & 0.319 & $+0.014 -0.013$ & $+0.016 -0.035$ \\
\hline
\end{tabular}
\caption{Highest and lowest fit masses for the vector meson with
degenerate quarks, based on an analysis of the fit regions 13--23
to 18--23.  We take the difference between the highest (lowest)
mass and the best-fit mass as a measure of the systematic error.}
\label{tab:V_stability}
\end{center}
\end{table}
 %

\begin{figure}[htbp]
\begin{center}
\leavevmode
\epsfysize=300pt
  \epsfbox[20 30 620 600]{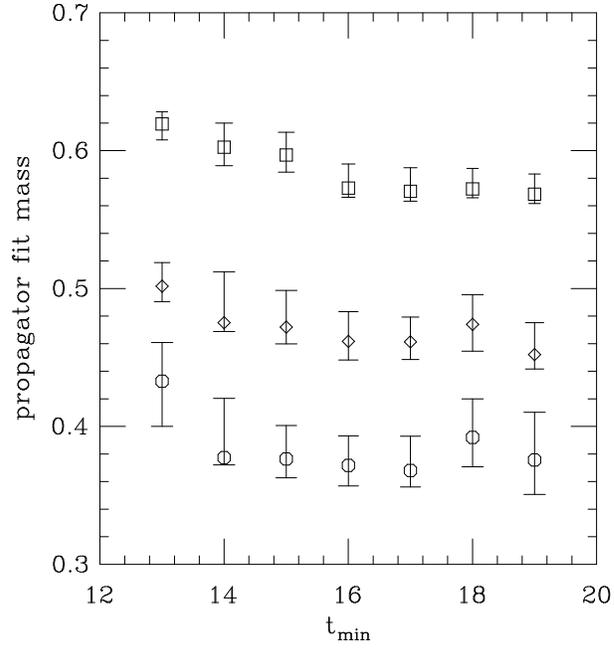}
\end{center}

\begin{center}
\leavevmode
\epsfysize=300pt
  \epsfbox[20 30 620 600]{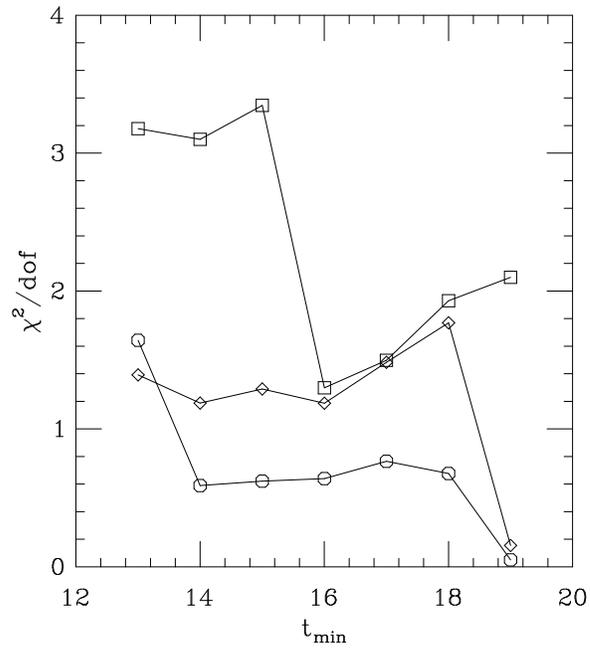}
\end{center}
\caption{Fit stability plots for the nucleon.  The top graph shows
the variation of fit mass with $t_{\rm min}$, fixing $t_{\rm
max}=22$.  The bottom graph shows the corresponding $\chi^2/$dof.
The different symbols refer to different $\kappa$ values:
squares~=~0.14144, diamonds~=~0.14226, circles~=~0.14262.}
\label{fig:N_stability}
\end{figure}

In \fig{fig:N_stability} and \tab{tab:N_stability} we show the
corresponding data for the nucleon.  These provide convincing
evidence of plateaux at all three $\kappa$ values, beginning at
$t_{\rm min}=16$ for the heaviest and at $t_{\rm min}=14$ for the
two lightest masses.  We note that the large upper error bars in
the latter data at $t_{\rm min}=14$ and 15 may indicate the
sensitivity of the bootstrap sampling to the tail of the
excited-state contributions.  For this reason, we take $t_{\rm
min}=16$ for all three $\kappa$ values.  The systematic error
estimates in Table~\ref{tab:N_stability} are well within the
statistical errors, giving us confidence in our choice of fit
range.
 %
%
%
%
\begin{table}
\begin{center}
\begin{tabular}{|c||c|c|c|c|c|}
\hline
 & & & & \multicolumn{2}{|c|}{error} \\
\cline{5-6}
$\kappa$ & $m_N^{\rm high}$ & $m_N^{\rm low}$ & $m_N$ & statistical &
systematic \\
\hline
0.14144  & 0.573 & 0.568 & 0.573 & $+0.015 -0.007$ & $+0.000 -0.005$ \\
0.14226  & 0.475 & 0.452 & 0.462 & $+0.020 -0.014$ & $+0.013 -0.010$ \\
0.14262  & 0.392 & 0.372 & 0.372 & $+0.024 -0.016$ & $+0.020 -0.000$ \\
\hline
\end{tabular}
\caption{Highest and lowest fit masses for the nucleon, based on
an analysis of the fit regions 16--22 to 19--22 for the heaviest
and 14--22 to 19--22 for the others.  We take the difference
between the highest (lowest) mass and the best-fit mass as a
measure of the systematic error.}
\label{tab:N_stability}
\end{center}
\end{table}
 %

We calculate the pseudoscalar decay constant from the ratio
\begin{equation}
\frac{\sum_{\bm{x}} \langle A_4(\bm{x},t)
P^{\dagger}(0)
\rangle}
{\sum_{\bm{x}} \langle P(\bm{x},t) P^{\dagger}(0)
\rangle} \sim \frac{f_{P} m_{P}}{Z_A\langle 0 | P | P
\rangle}
\tanh{m_{P} ( L_t/2 - t)},
\end{equation}
fitting to the same time slice range as in the fit to the
pseudoscalar correlator.  We use the parameters from this latter
fit to fix $m_{P}$ and the matrix element $\langle 0 | P | P
\rangle$.  We find that this particular ratio of correlators gives
the cleanest signal from which to extract
$f_{P}$~\cite{UKQCD_smearing}.  We determine $f_V$ by fitting to
\begin{equation}
\sum_{j = 1}^3 \sum_{\bm{x}} \langle V_j(\bm{x}, t)
V_j^\dagger (0)\rangle \sim \frac{3 m_V^3}{2 Z_V^2 f_V^2}
e^{-m_V L_t/2} \cosh m_V (L_t/2 - t).
\end{equation}
Here, $Z_A$ and $Z_V$ are the factors required to ensure that
the lattice currents obey the correct current algebra in the
continuum limit~\cite{bochicchio,pittori}.
 %
%



\section{Results}

\subsection{Masses and Decay Constants in Lattice Units}

In Table~\ref{tab:masses_decays} we present the masses and decay
constants of the vector and pseudoscalar mesons calculated from
both degenerate-quark and non-degenerate-quark correlators.  The
$\chi^2/$dof are all satisfactory, being generally between 0.5 and
2.  The degenerate-quark data may be compared with our results
from the first 18 configurations, presented
in~\cite{UKQCD_hadrons_npb}, where our fit range was 12 -- 16 for
all hadrons.  Our new estimates are within $1\sigma$ of the 18
configuration estimates.  The errors are reduced by a factor of
approximately 2.
 %
\begin{table}
\begin{center}
\begin{tabular}{|c||c|c|c||c|c|c|}
\hline
 & \multicolumn{3}{|c||}{degenerate} & \multicolumn{3}{|c|}{non-degenerate} \\
\hline
$\kappa_1$ & 0.14144 & 0.14226 & 0.14262 & 0.14144 & 0.14262 & 0.14226 \\
$\kappa_2$ & 0.14144 & 0.14226 & 0.14262 & 0.14226 & 0.14144 & 0.14262 \\
\hline
\hline
$m_{P}$   & 0.298\er{2}{2} & 0.214\er{2}{3} & 0.167\er{3}{4}
           & 0.259\er{2}{2} & 0.241\er{2}{3} & 0.192\er{3}{3} \\
$\chi^2/{\rm dof}$
           & 8.7/7   & 6.9/7   & 7.2/7   & 6.3/7   & 5.3/7   & 6.9/7 \\
\hline
$m_V$      & 0.389\er{7}{6}  & 0.343\er{9}{7} & 0.319\err{14}{13}
           & 0.370\er{6}{5}  & 0.360\er{8}{6} & 0.331\err{11}{10} \\
$\chi^2/{\rm dof}$
           & 13/7    & 7.8/9   & 4.0/9   & 12/9   & 9.1/9   & 5.4/9  \\
\hline
$m_V^2-m_{P}^2$
           & 0.063\er{5}{4} & 0.072\er{6}{5} & 0.074\er{9}{8}
           & 0.070\er{5}{3} & 0.071\er{6}{4} & 0.073\er{7}{6} \\
\hline
\hline
$f_{P}/Z_A$   & 0.0624\err{7}{13} & 0.0512\err{6}{15} & 0.0452\err{8}{21}
                 & 0.0567\err{7}{13} & 0.0539\err{6}{15} & 0.0482\err{7}{17} \\
$\chi^2/{\rm dof}$
           & 12/8   & 9.2/8   & 8.3/8   & 11/8   & 9.2/8   & 8.1/8 \\
\hline
$1/(f_V Z_V)$  & 0.314\er{8}{7}  & 0.345\er{8}{9}  & 0.356\err{9}{17}
                 & 0.332\er{6}{7}  & 0.336\er{8}{9}  & 0.350\err{9}{12} \\
$\chi^2/{\rm dof}$
           & 13/7    & 7.8/9   & 4.0/9   & 12/9   & 9.1/9   & 5.4/9  \\
\hline
\end{tabular}
\caption{{Masses and decay constants in lattice units of mesons
composed of degenerate and non-degenerate quarks.}\label{tab:masses_decays}}
\end{center}
\end{table}
 %

\begin{figure}[htbp]
\begin{center}
\leavevmode
\epsfysize=300pt
  \epsfbox[20 30 620 600]{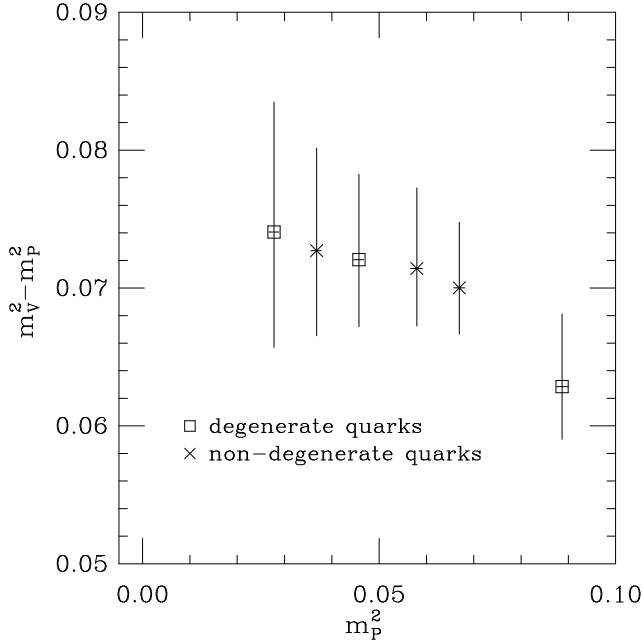}
\end{center}
\caption{Vector-pseudoscalar meson mass splitting for both degenerate-
and non-degenerate-quark data.} \label{fig:m2V_m2P}
\end{figure}

As noted previously~\cite{UKQCD_hadrons_lett,UKQCD_hadrons_npb},
experimental data suggests that the hyperfine splitting,
$m_V^2-m_{P}^2$, should be only weakly dependent on the
quark masses for light hadrons, and this we observed
within large errors.  Our higher-statistics results quoted in
Table~\ref{tab:masses_decays}, are shown in
Figure~\ref{fig:m2V_m2P}.  The new data for both degenerate and
non-degenerate quarks is entirely consistent with our earlier
results, but with significantly smaller errors.
 %
 %

In Table ~\ref{tab:baryons} we present the masses of the nucleon
and $\Delta$ computed using degenerate quarks only.  Again, the
$\chi^2/$dof of the fits is acceptable.  Our estimates for the
nucleon mass are about $2\sigma$ below our estimates based on the
first 18 configurations~\cite{UKQCD_hadrons_npb}, with only
slightly smaller errors.  It is clear from
Figure~\ref{fig:N_stability} that $t_{\rm min}=12$ is not in the
plateau region when using 60 configurations.  We have used the
increase in sample size primarily to reduce the systematic error
in the mass due to contamination by excited states.  However, this
may not be the only effect.  We observe increasing fluctuations in
the nucleon mass with decreasing quark mass, and fluctuations
which decrease the nucleon mass tend to dominate the statistical
average.  The influence of such fluctuations in the present
analysis may be more pronounced because we fit the nucleon
correlator further from the source.
 %
\begin{table}
\begin{center}
\begin{tabular}{|c||c|c|c|}
\hline
& \multicolumn{3}{|c|}{degenerate} \\
\hline
$\kappa$   & 0.14144          & 0.14226           & 0.14262  \\
\hline
\hline
$m_N$      & 0.573\err{15}{6} & 0.462\err{20}{14} & 0.372\err{24}{16} \\
$\chi^2/{\rm dof}$
           & 6.5/5            & 5.9/5             & 3.2/5 \\
\hline
$m_\Delta$ & 0.646\err{12}{10}& 0.577\err{31}{21} & 0.556\err{66}{46} \\
$\chi^2/{\rm dof}$
           & 0.7/4            & 1.1/4             & 1.7/4 \\
\hline
\end{tabular}
\caption{{Masses in lattice units of baryons composed of
degenerate quarks.}\label{tab:baryons}}
\end{center}
\end{table}
 %

Apart from at the highest quark mass, where our new estimate is
$3\sigma$ lower, our estimates for $m_\Delta$ agree with our
previous results. The errors have not decreased, probably
because of the extended fitting range.

We have looked for evidence of correlations between successive
configurations, by varying the bin size in a jackknife error
analysis.  This did not reveal any significant effects in the
hadron time slice correlators.

\begin{figure}[htbp]
\begin{center}
\leavevmode
\epsfysize=300pt
  \epsfbox[20 30 620 600]{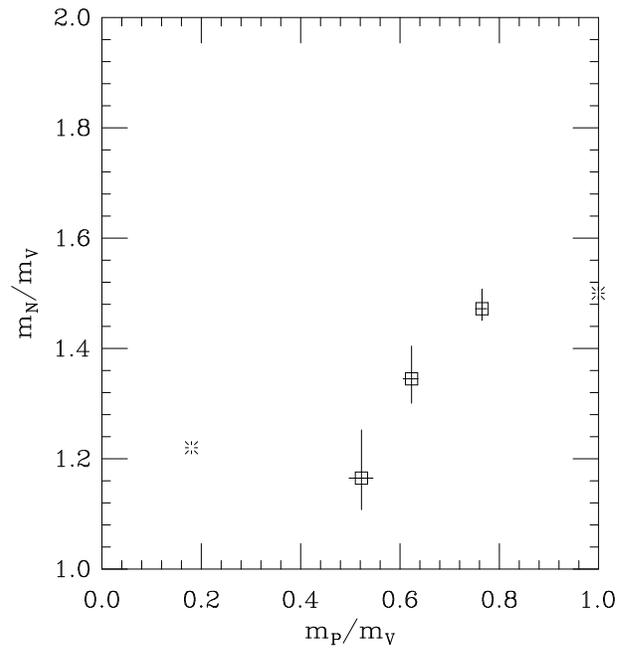}
\end{center}
\caption{Edinburgh plot for degenerate-quark data.}
\label{fig:edinburgh}
\end{figure}

In Figure~\ref{fig:edinburgh} we show the Edinburgh plot for our
degenerate-quark data.  The corresponding mass ratios are
given in Table~\ref{tab:ratios}.  The apparently alarming fall of
the data points may not indicate any discrepancy with experiment,
as we shall see that the chirally-extrapolated value of $m_N/m_V$
is 1.07\err{11}{8}, within $2\sigma$ of the experimental value, as
may be deduced from Table~\ref{tab:phys_baryons}.  Indeed, our
higher-statistics results are only $1-2\sigma$ below our earlier
result using local sources and sinks~\cite{UKQCD_hadrons_lett}.
We attribute this difference to our new, lower estimates for the
nucleon mass.  We remark that were the systematic errors in the
vector meson mass, discussed above, to be included somehow, the
tendency would be for the upper and right-hand error bars to
increase.
 %
 %
\begin{table}
\begin{center}
\begin{tabular}{|c||c|c|}
\hline
$\kappa$   & $m_{P}/m_V$  & $m_N/m_V$\\\hline
0.14144    & 0.77\er{1}{1} & 1.47\er{4}{2}\\
0.14226    & 0.62\er{1}{2} & 1.35\er{6}{4}\\
0.14262    & 0.52\er{2}{2} & 1.17\er{9}{6}\\
\hline
\end{tabular}
\caption{{Mass ratios from degenerate-quark data, used in the
Edinburgh plot.}\label{tab:ratios}}
\end{center}
\end{table}
 %



\subsection{Quark-Mass Dependences}

We fit our lattice estimates for the meson masses, decay constants
and their ratios, for all six quark-mass combinations, to the
expression~(\ref{eq:ndg_fit_form}) with $a_3$, the coefficient of
$|m_2 - m_1|$, (A) unconstrained and (B) constrained to be zero.
In \tab{tab:fit_AB} we present values for the fit coefficients.
For all the quantities, the values obtained for $a_3$ in fit (A)
are consistent with zero to within one or two standard deviations.
The $\chi^2/$dof for all the fits are satisfactory, although
mostly a little larger for fit (B), and in each case the two fits
give completely consistent values for $a_1$ and $a_2$.  This is
the numerical evidence that these physical quantities depend only
on the sum of the quark masses.  As there is no theoretical
justification for $a_3\neq 0$, nor any support for this from our
data, from hereon we use only the results of fit (B).
 %
\begin{table}
\begin{center}
\begin{tabular}{|c||c|c|c|c|}
\hline
 & \multicolumn{3}{|c|}{fit (A) parameters} & \\
\cline{2-4}
 & \makebox[5em][c]{$a_1$} & \makebox[5em][c]{$a_2$}
 & \makebox[5em][c]{$a_3$} & $\chi^2/$dof \\
\hline
$m_{P}^2$               & 0.0       & 2.12\er{4}{3}
                        & $-0.01$\er{1}{1}    & 4.3/3 \\
\hline
$m_V$                   & 0.29\er{2}{1}       & 2.3\er{4}{3}
                        & 0.1\er{1}{1}        & 1.9/3 \\
\hline
$f_{P}/Z_A$             & 0.041\er{1}{2}      & 0.53\er{3}{2}
                        & $-0.01$\er{1}{1}    & 4.8/3 \\
\hline
$1/(f_V Z_V)$           & 0.38\er{1}{2}       & $-1.4$\er{4}{2}
                        & 0.0\er{1}{1}        & 1.2/3 \\
\hline
$f_{P}/(m_V Z_A)$       & 0.142\er{6}{9}      & 0.5\er{2}{2}
                        & $-0.10$\er{6}{7}    & 3.1/3 \\
\hline
\hline
 & \multicolumn{3}{|c|}{fit (B) parameters} & \\
\cline{2-4}
 & \makebox[5em][c]{$a_1$} & \makebox[5em][c]{$a_2$}
 & \makebox[5em][c]{$a_3$} & $\chi^2/$dof \\
\hline
$m_{P}^2$               & 0.0                & 2.12\er{4}{3}
                        & 0.0                & 9.5/4 \\
\hline
$m_V$                   & 0.29\er{1}{1}      & 2.5\er{3}{3}
                        & 0.0                & 2.5/4 \\
\hline
$f_{P}/Z_A$             & 0.040\er{1}{2}     & 0.53\er{3}{2}
                        & 0.0                & 7.1/4 \\
\hline
$1/(f_V Z_V)$           & 0.38\er{1}{2}      & $-1.4$\er{3}{2}
                        & 0.0                & 1.3/4 \\
\hline
$f_{P}/(m_V Z_A)$       & 0.142\er{6}{9}     & 0.4\er{2}{2}
                        & 0.0                & 5.6/4 \\
\hline
\end{tabular}
\caption{{Fit parameters for masses, decay constants and ratios,
using the fit form described in
equation~(\protect\ref{eq:ndg_fit_form}), (A) with $a_3$
unconstrained, and (B) with $a_3=0$.}
\label{tab:fit_AB}}
\end{center}
\end{table}
 %

\subsection{Chiral Extrapolations}

\begin{figure}[htbp]
\begin{center}
\leavevmode
\epsfysize=300pt
  \epsfbox[20 30 620 600]{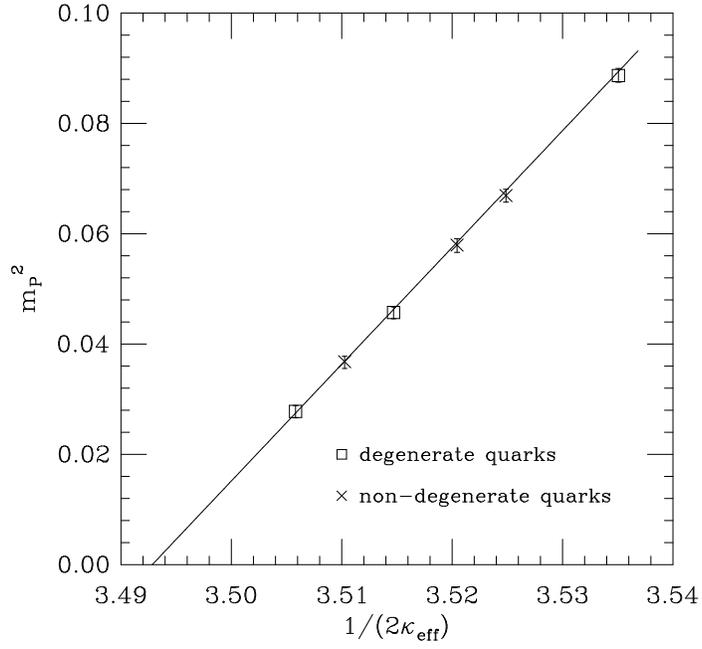}

\leavevmode
\epsfysize=300pt
  \epsfbox[20 30 620 600]{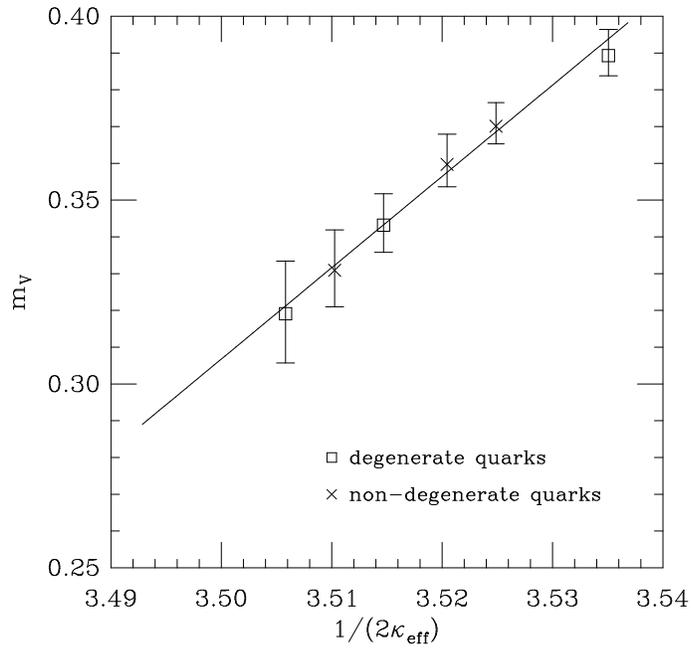}
\end{center}
\caption{Chiral extrapolations of the pseudoscalar and vector
meson masses.}
\label{fig:chiral_meson}
\end{figure}

Firstly, we fit our estimates of the pseudoscalar meson mass for
all six quark-mass combinations to the form in
\eqn{eq:dg_two_kappas}, and obtain \kcrit\ from extrapolating the
fit in both $\kappa$'s to $m_{P}^2(\kcrit,\kcrit) = 0$.  This
gives
\begin{equation}
\kcrit = 0.14315\mbox{\er{2}{2}},
\end{equation}
in good agreement with, although significantly more accurate than,
our estimate from 18 configurations~\cite{UKQCD_hadrons_npb}.  In
\fig{fig:chiral_meson}, we present the plot of $m^2_{P}$ versus
$1/2\kappa_{\rm eff}$ from which we derive \kcrit, defining an
effective $\kappa$ as
\begin{equation}
\frac{1}{\kappa_{\rm eff}} =
	\left( \frac{1}{2 \kappa_1} + \frac{1}{2 \kappa_2} \right).
\label{eq:eff_kappa}
\end{equation}
The fact that both the degenerate- and non-degenerate-quark data
agree well with the fit is graphical evidence of our claim in the
previous section that the pseudoscalar meson mass depends only on
the sum of the quark masses, and confirms the observation of
reference~\cite{omero}.  The chiral extrapolation of $m_V$ to
\kcrit\ using \eqn{eq:dgV_two_kappas} is also shown in
\fig{fig:chiral_meson}; again, we conclude from the good agreement
between the data and the fit that $m_V$ depends only on the sum of
the quark masses.
 %
 %

\begin{figure}[htbp]
\begin{center}
\leavevmode
\epsfysize=300pt
  \epsfbox[20 30 620 600]{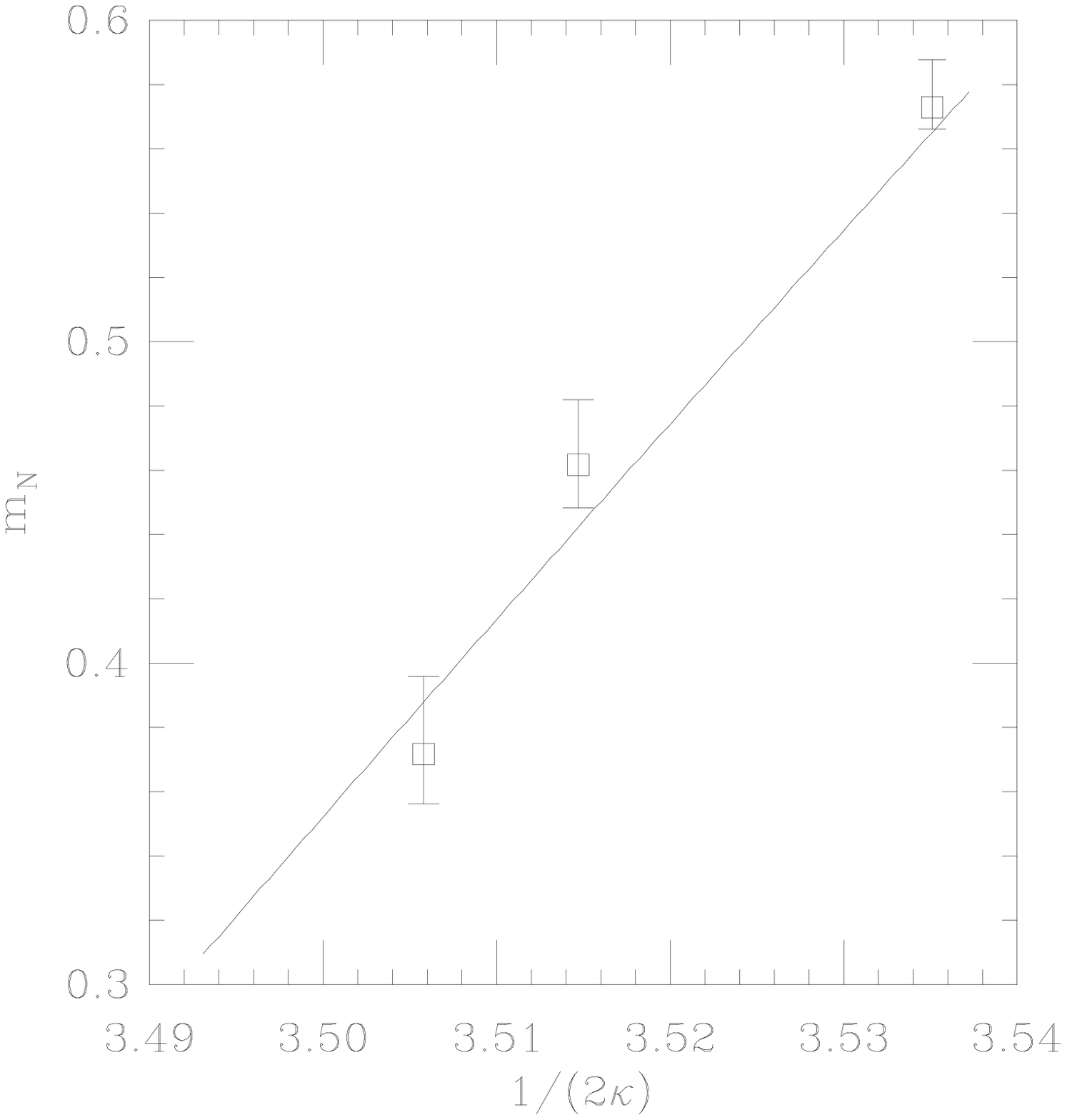}

\leavevmode
\epsfysize=300pt
  \epsfbox[20 30 620 600]{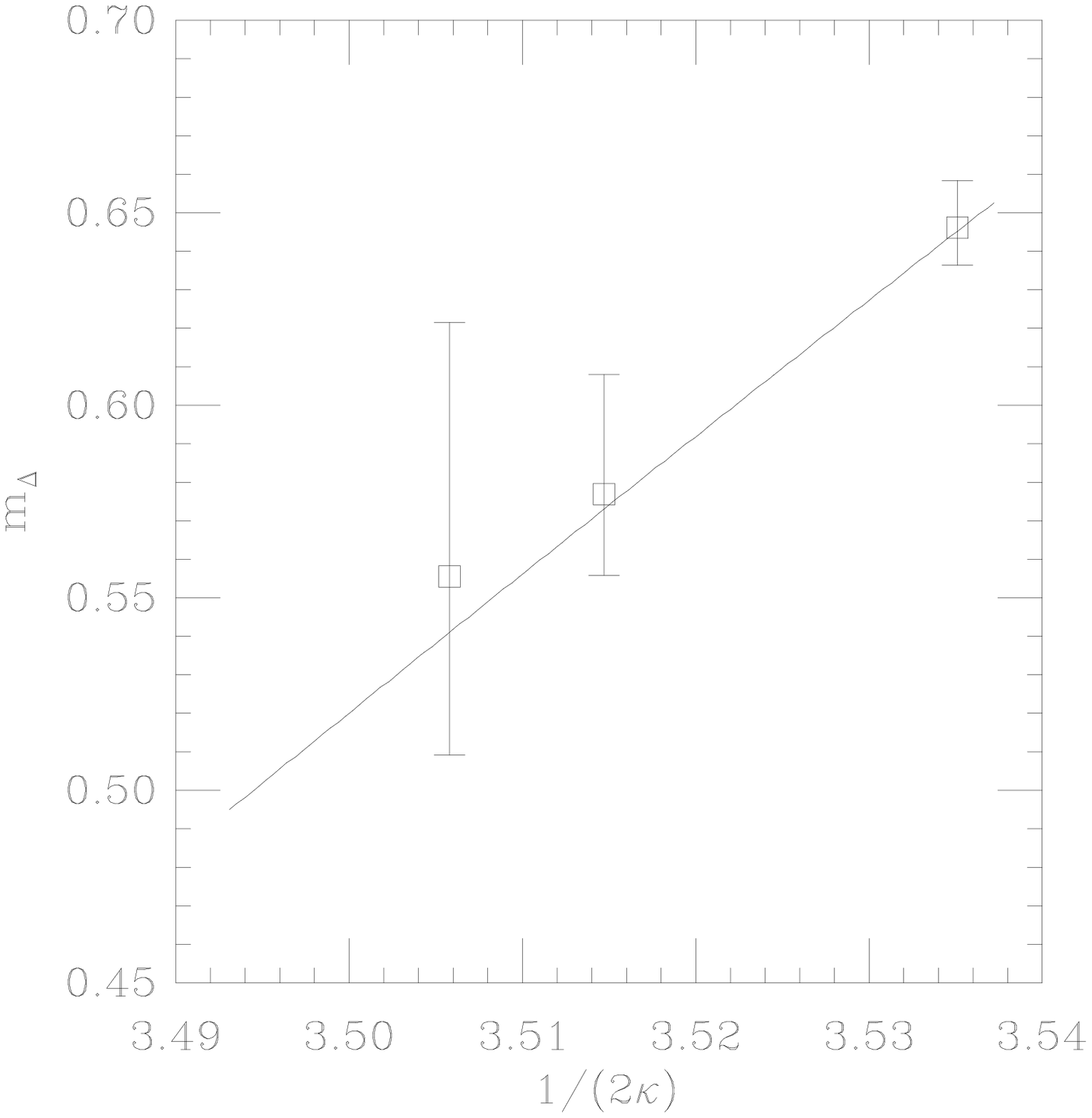}
\end{center}
\caption{Chiral extrapolations of $m_N$ and $m_\Delta$ for
degenerate-quark data only.}
\label{fig:chiral_baryon}
\end{figure}

We present, in \fig{fig:chiral_baryon}, the chiral extrapolations
of the nucleon and $\Delta$.  It is evident that for the nucleon
the quality of the linear fit is rather poor.  This is supported
by the fact that the $\chi^2/$dof for this fit is 5.1, compared to
the value 2.4 obtained for the pseudoscalar meson fit and 0.63
obtained for the vector meson fit.  Negative curvature of the
nucleon has been observed before (see for
example~\cite{loft_degrand,ape6.0}).  A linear fit to our data for
$m_N^2$ gives a smaller $\chi^2/$dof of 2.7, as well as a
substantially smaller estimate for the nucleon mass in the chiral
limit.  However, because there is no theoretical justification for
this choice of extrapolation, and because we cannot reliably
compare different choices with only three data points, we quote
results only from the linear extrapolation of $m_N$, noting that
there is significant uncertainty in this procedure.  We see no
such problem with the chiral extrapolation of $m_\Delta$, although
the errors are larger.
 %
 %

The inverse lattice spacing, $a^{-1}$, in physical units, obtained
from each of the $\rho$, nucleon and $\Delta$ masses, is given in
Table~\ref{tab:scales}.  Throughout the remainder of this paper,
we will use $m_\rho$ to set the scale.  In so far as there is good
agreement between the scales from the string tension and $m_\rho$,
our results in physical units are not especially dependent on the
chiral extrapolation.  The corresponding values for the nucleon
and $\Delta$ masses in physical units are given in
Table~\ref{tab:phys_baryons}.
 %
 %
\begin{table}
\begin{center}
\begin{tabular}{|c|c|c|}\hline
physical quantity & lattice value     & $a^{-1}$~(GeV)     \\\hline
$m_\rho$          & 0.29\er{1}{1} & 2.7\er{1}{1} \\
$m_N$             & 0.31\er{3}{2} & 3.0\er{2}{3} \\
$m_\Delta$        & 0.50\er{4}{3} & 2.5\err{2}{2} \\\hline
$\sqrt{K}$        & 0.161(3)          & 2.73(5)\\\hline
\end{tabular}
\end{center}
\caption{Chirally-extrapolated lattice masses and corresponding
scales compared with the scale from the string
tension~\protect\cite{UKQCD_hadrons_npb}.}
\label{tab:scales}
\end{table}
 %
 %


\subsection{Determination of the Strange Quark Mass}

We calculate the value of \kstrange\ from the fit to the ratio
$m_{P}^2(\kappa_1,\kappa_2)/m_\rho^2$ for all six quark-mass
combinations, extrapolating $\kappa_1$ to \kcrit, and using
$\kappa_2$ to match the ratio to its experimental value,
$m_K^2/m_\rho^2 = 0.413$, giving
\begin{equation}
\kappa_2 = \kstrange = 0.1419\mbox{\er{1}{1}.}
\label{eq:kstrange}
\end{equation}
This value agrees well with that obtained from degenerate-quark
data with smeared sinks in reference~\cite{UKQCD_hadrons_npb}.  We
note that $m_s$ lies between two of our quark masses, so that our
strange-quark results are obtained by interpolation, an
intrinsically more robust procedure than extrapolation.

Our value of $\kappa_s$ corresponds, in the lattice regularisation, to
a strange quark mass in physical units of
\beq
m_s(a)=\frac{1}{2a}
\left( \frac{1}{\kappa_s}-\frac{1}{\kappa_{{\rm crit}}} \right)
= 82\er{8}{8}~{\rm MeV},
\label{eq:ms}
\eeq
where we have used $m_\rho$ to determine the lattice spacing in
physical units.  From this we can determine the renormalised
strange quark mass in the $\overline{{\rm MS}}$ scheme at a
reference renormalisation scale of $\mu=2$~GeV:
\beq
m_s^{\msbar}(2\,\gev) = 109\err{11}{11},
\label{eq:msmsbar}
\eeq
where we have used the perturbative value of the renormalisation
constant relating $m_s(a)$ and $m_s^{\msbar}(\mu )$~\cite{gmphs},
together with the effective coupling defined in reference~\cite{LM_91}.
This result agrees with the conclusion from previous lattice
evaluations of the strange quark mass (see reference~\cite{abada}
for a simulation with Wilson fermions and reference~\cite{mssv}
for one with the clover action at $\beta=6.0$), that $m_s^{\msbar}
(2\,\gev )$ is at the lower end of expectations \cite{pdg,dd}.
The value given in Equation~(\ref{eq:msmsbar}) can be compared to
$100 \pm 6$~MeV quoted in reference~\cite{abada} and $89 \pm
9$~MeV quoted in reference~\cite{mssv}.


\subsection{Spectrum and Decay Constants in Physical Units}

We present the results for meson masses and decay constants,
extrapolated/interpolated to the physical $\kappa$ values, in
\tab{tab:extraps}, using the parameters of fit (B).  As indicated
in \tab{tab:extraps}, the ratios of the decay constants need to be
multiplied by renormalisation constants before they can be
compared with experimental numbers.  The perturbative estimates
for these renormalisation constants~\cite{pittori}, using the
effective coupling, are
\begin{equation}
Z_A \simeq 0.97, \qquad Z_V \simeq 0.83.
\label{eq:f_renorm_consts}
\end{equation}
Incorporating these values for the renormalisation constants and
using the lattice scale from $m_\rho$ in Table~\ref{tab:scales},
we obtain the meson masses and decay constants in physical units
presented in \tab{tab:physicals}.
 %
\begin{table}
\begin{center}
\begin{tabular}{|c||c|}
\hline
$m_{\eta_s}$           & 0.251\err{13}{11}  \\
$m_{K^*}$              & 0.326\err{13}{12}  \\
$m_\phi$               & 0.364\err{12}{11}  \\
$m_{K^*}^2 - m_K^2$    & 0.075\er{6}{5}     \\\hline
$f_\pi/Z_A$            & 0.040\er{1}{2}     \\
$f_K/Z_A$              & 0.048\er{1}{2}     \\
$f_K/f_\pi$            & 1.20\er{3}{2}      \\\hline
$1/(f_\rho Z_V)$       & 0.380\err{9}{16}   \\
$1/(f_{K^*} Z_V)$      & 0.359\err{6}{11}   \\
$1/(f_\phi Z_V)$       & 0.337\er{4}{7}     \\
$f_\phi/f_\rho$        & 1.13\er{2}{3}      \\
$f_{K^*}/f_\rho$       & 1.06\er{1}{2}      \\\hline
$f_\pi/(m_\rho Z_A)$   & 0.142\er{6}{9}     \\
$f_K/(m_\rho Z_A)$     & 0.165\er{7}{8}     \\
$f_K/(m_{K^*} Z_A)$    & 0.148\er{4}{6}     \\
\hline
\end{tabular}
\caption{{Extrapolated/interpolated values of meson masses and
decay constants in lattice units, calculated using the fit (B)
parameters in Table~\protect\ref{tab:fit_AB}.}
\label{tab:extraps}}
\end{center}
\end{table}
 %
\begin{table}
\begin{center}
\begin{tabular}{|c||c|c|}
\hline
 & lattice estimates & experiment \\
\hline
$m_{\eta_s}$      & 670\err{10}{10} MeV  &``686 MeV''\\
$m_{K^*}$         & 868\er{9}{8} MeV     & 892 MeV\\
$m_\phi$          & 970\err{20}{10} MeV  & 1020 MeV\\
$m_{K^*}^2-m_K^2$ & 0.53\er{1}{1} (GeV)$^2$   & 0.55 (GeV)$^2$\\
\hline
$f_\pi$           & 102\er{6}{7} MeV     & 132 MeV \\
$f_K$             & 123\er{5}{6} MeV     & 160 MeV \\
\hline
$1/f_\rho$        & 0.316\err{7}{13}     & 0.28 \\
$1/f_{K^*}$       & 0.298\er{5}{9}       &  \\
$1/f_\phi$        & 0.280\er{3}{6}       & 0.23 \\
\hline
$f_\pi/m_\rho$    & 0.138\er{6}{9}       & 0.172 \\
$f_K/m_\rho$      & 0.160\er{7}{8}       & 0.208 \\
$f_K/m_{K^*}$     & 0.144\er{4}{6}       & 0.179 \\
\hline
\end{tabular}
\caption{Values of meson masses and decay constants in physical
units, using the scale from $m_\rho$.}
\label{tab:physicals}
\end{center}
\end{table}
 %

The results for $m_{K^*}$, $m_{K^*}^2-m_K^2$ and $m_\phi$,
although $2-3\sigma$ below experiment, provide support for our
determination of $\kappa_s$. However, they are open to the
interpretation that the vector meson masses may be slightly
underestimated relative to the pseudoscalar meson masses, as a
result of the suppression of spin-splittings in the quenched
approximation.  Our result for $m_{K^\ast}$ is significantly more
precise than that obtained with the standard Wilson action at
$\beta=6.0$ by Lipps et al.~\cite{lipps} who quote $930(40)$~MeV,
and by Loft \& DeGrand~\cite{loft_degrand} who obtain
$761(122)$~MeV.  It is comparable with the value 896(17)~MeV
obtained using a renormalisation-group-improved action by
Iwasaki~\cite{iwasaki}.  Lipps et al.~and Iwasaki both use
$m_\phi$ to determine $\kappa_s$, based on the assumption that the
$\phi$ is pure $s\bar{s}$, and so we are only able to compare our
result for $m_\phi$ with that of Loft \& DeGrand, who give
868(114)~MeV.  Loft \& DeGrand determine $\kappa_s$ from the
$\eta_s$, which is the hypothetical pure $s\bar{s}$ pseudoscalar
meson, whose theoretically-expected mass of 686~MeV~\cite{lipps}
agrees well with our calculation.

Martinelli and Maiani~\cite{MM_86} noted that the ratio
$(m_{K^\ast}-m_\rho)/(m_K^2-m_\pi^2)$ typically gives a value for
the inverse lattice spacing which is lower than that obtained
using other physical quantities.  The ratio is
estimated by assuming that $m_V$ and $m_{P}^2$ are linear in the
sum of the quark masses, Equations~(\ref{eq:dg_two_kappas}) and
(\ref{eq:dgV_two_kappas}), so that:
\begin{equation}
m_V(\kappa_1,\kappa_2) = a_V + \frac{b_V}{b_{P}}
m_{P}^2(\kappa_1,\kappa_2)
\end{equation}
and
\begin{equation}
\frac{m_{K^\ast} - m_\rho}{m_K^2 - m_\pi^2} =
\frac{m_V(\kappa_s,\kcrit)   - m_V(\kcrit,\kcrit)}
{m_{P}^2(\kappa_s,\kcrit) - m_{P}^2(\kcrit,\kcrit)} = \frac{b_V}{b_{P}}.
\end{equation}
We obtain $2.3\err{3}{3}$~GeV, in good agreement with our
low-statistics results~\cite{UKQCD_hadrons_npb}.  This is
consistent with the scales we obtain from other physical
quantities, although it remains on the low side.

Our results for the pseudoscalar decay constants are $5 - 7\sigma$
below their experimental values.  This has been noted recently in
simulations using the standard Wilson
action~\cite{weingarten_decay}.  These authors argue that a
smaller value of the decay constant is to be expected in the
quenched approximation than in the full theory, as a consequence
of the smaller wavefunction at the origin.  The discrepancy may
also be partly due to our use of the perturbative value for $Z_A$.
Recent non-perturbative calculations at $\beta=6.0$, and at a
single value of the quark mass, put the value of $Z_A \sim
1.09(3)$~\cite{non-pert_ren}, somewhat higher than the
perturbative value.  Thus, it may be hoped that a full
non-perturbative evaluation of $Z_A$ at $\beta=6.2$ will raise our
estimates of the decay constants, bringing them closer to the
experimental values.  The uncertainty in the renormalisation
constant is removed in the ratio $f_K/f_\pi = 1.20$\er{3}{2},
which agrees well with the experimental value of 1.22.
This suggests that, although we are working in the quenched
approximation, we obtain correctly the dependence on the strange
quark mass.  We cannot attribute the problem with the overall
normalisation to quenching until we have a precise
non-perturbative determination of the axial current
renormalisation.

\begin{figure}[htbp]
\begin{center}
\leavevmode
\epsfysize=300pt
  \epsfbox[20 30 620 600]{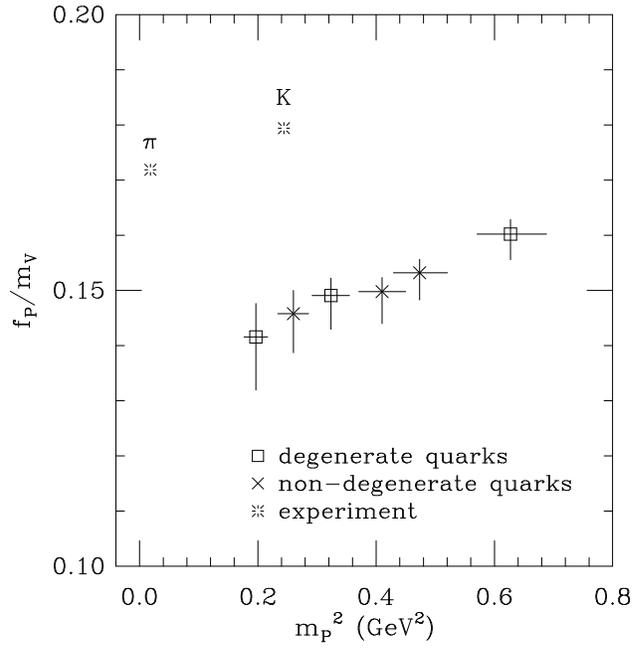}
\end{center}
\caption{$f_{P}/m_V$ against $m_{P}^2$ in physical units using
the scale from $m_\rho$.  Experimental points for the pion and
kaon are also shown.}
\label{fig:fP_mV_vs_m2P}
\end{figure}

Experimentally, the ratio of the pseudoscalar decay constant to
the vector meson mass is fairly insensitive to the
$SU(3)$-flavour-symmetry breaking.  In \fig{fig:fP_mV_vs_m2P} we
plot this ratio versus $m_{P}^2$ for both the degenerate- and
non-degenerate-quark data.  The slope agrees well with that of the
experimental data and we note that an increase in $Z_A$ of order
25\% would give excellent agreement between the two.
 %
 %

\begin{figure}[htbp]
\begin{center}
\leavevmode
\epsfysize=300pt
  \epsfbox[20 30 620 600]{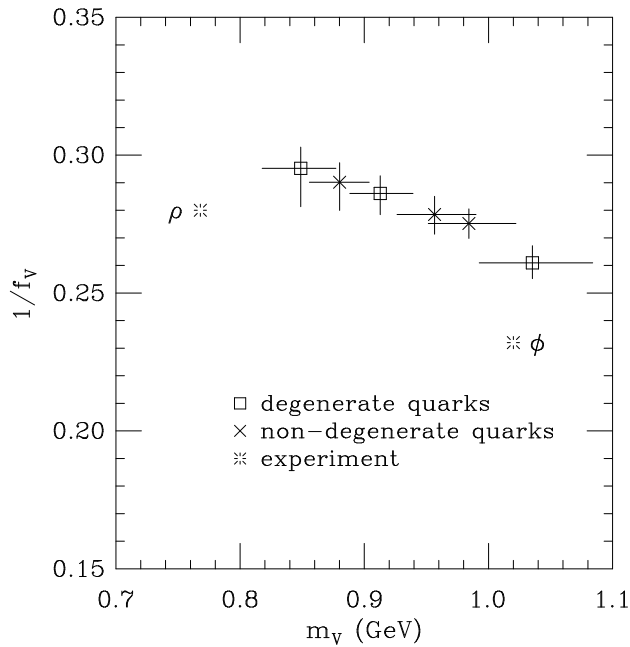}
\end{center}
\caption{$1/f_V$ against $m_V$ in physical units using the scale
from $m_\rho$.  Experimental points for the $\rho$ and $\phi$
mesons are also shown.}
\label{fig:fV_vs_mV}
\end{figure}

Our estimates for the vector meson decay constants, $1/f_\rho$ and
$1/f_\phi$, given in Table~\ref{tab:physicals}, lie above the
experimental values.  This is shown for all our data in
\fig{fig:fV_vs_mV}.  The discrepancy is small for the $\rho$, but
it becomes significant for the $\phi$, although the slope of the
data is consistent with experiment.  The sign of the discrepancy
is opposite to that expected from the suppression of the
wavefunction at the origin by the quenched approximation,
suggesting that other effects are important.  The difference is
less likely than in the case of the pseudoscalar decay constant to
be due to our use of the perturbative value of the renormalisation
constant, because at least at $\beta=6.0$ $Z_V$ is known to be close
to its non-perturbative value~\cite{non-pert_ren}.  Thus, we
suspect that significant discretisation and/or finite-volume
errors are present.
 %
 %

For the baryon data in \tab{tab:baryons}, we perform
extrapolations and interpolations in the single $\kappa$ value to
obtain our predictions for the physical masses.  The $\Delta$ and
$\Omega$ baryons both have $J^{P} = \frac{3}{2}^+$, so by
interpolating our $\Delta$ fits to \kstrange\ we can obtain
$m_\Omega$, as shown in \tab{tab:phys_baryons}.  We are encouraged
by the good correspondence between our computed values and the
experimental masses presented in the table.  However, the low
confidence we have in the linear chiral extrapolation of the
nucleon means that we cannot attach much significance to the
comparison between the computed and experimental values for $m_N$.
The extrapolation of $m_\Delta$ does not have this problem, and
the interpolation to \kstrange\ gives a value of $m_\Omega$ very
close to the experimental number.  This indicates that our
calculation of \kstrange\ can be applied sensibly to the baryon
sector.  Our value for $m_\Omega$ is to be compared with that of
Lipps et al.~\cite{lipps} who quote $1650(150)$~MeV, Loft \&
DeGrand~\cite{loft_degrand} who quote $1512(144)$~MeV, and
Iwasaki~\cite{iwasaki} who quotes $1737(77)$~MeV.
 %
\begin{table}
\begin{center}
\begin{tabular}{|c|c|c|c|} \hline
               & \multicolumn{2}{|c|}{extrapolated values} & experiment
\\
\cline{2-3}
               & lattice units     & MeV & \\
\hline
$m_N$          & 0.31\er{3}{2}  & 820\err{90}{60}      & 938 MeV
\\
$m_\Delta$     & 0.50\er{4}{3}  & 1300\errr{100}{100}  & 1232
MeV\\
$m_\Omega$     & 0.62\er{2}{2}  & 1650\err{70}{50}     & 1672
MeV \\
\hline
$m_\Omega/m_N$ & \multicolumn{2}{|c|}{2.0\er{2}{2}}   & 1.78 \\
\hline
\end{tabular}
\caption{Physical values of the baryon masses using $m_\rho$ to
set the scale, including the ratio $m_\Omega/m_N$.}
\label{tab:phys_baryons}
\end{center}
\end{table}
%

Our determination of the spin-$\frac{3}{2}$ baryon mass provides
us with an alternative means of determining $\kappa_s$, by
interpolating our lattice data to the physical value of
$m_{\Omega}$.  Using $m_\rho$ to set the scale, we obtain in this
way $\kappa_s = 0.1417\er{4}{3}$, $m_{\eta_s} =
740\err{50}{80}$~MeV and $m_\phi = 1010\err{20}{40}$~MeV.
Although there remains an implicit dependence of these estimates
on the chiral extrapolation, through our use of $m_\rho$, this can
be avoided by, for example, taking the scale from the string
tension.  It is evident from Table~\ref{tab:scales} that this
would give similar values.  Thus, this method of determining
$\kappa_s$ depends only on the mild assumption that the baryon
mass varies smoothly with quark mass close to the strange quark
mass.  The agreement, within the somewhat larger statistical
errors, with the value of $\kappa_s$ obtained from $m_K$, in
Equation~(\ref{eq:kstrange}), and with the corresponding mass
estimates in Table~\ref{tab:physicals}, provides a further check
on our determination of $\kappa_s$.  Finally, we note that
assuming that the $\phi$ is pure $s\bar{s}$ does not in practice
permit the determination of $\kappa_s$ from hadrons composed
solely of $s$ quarks, because the numerical data for the ratio
$m_\Omega/m_V(\kappa,\kappa)$ is only weakly dependent on
$\kappa$.



\section{Conclusions}
Despite working in the quenched approximation, our simulations
using the clover action at $\beta=6.2$ demonstrate good agreement
for meson masses and decay constants with the dependence on light
quark masses, including $SU(3)$-flavour-symmetry breaking,
expected from chiral perturbation theory. In particular, we find
evidence from simulations with non-degenerate quarks that the
dependences on the bare quark masses,
\begin{eqnarray}
m_{P}^2(\kappa_1,\kappa_2) & = & b_{P}\left( \frac{1}{2\kappa_1}
                                +\frac{1}{2\kappa_2}
                                -\frac{1}{\kcrit}\right)\\
m_V(\kappa_1,\kappa_2) & = & a_V + b_V\left( \frac{1}{2\kappa_1}
                                +\frac{1}{2\kappa_2}
                                -\frac{1}{\kcrit}\right),
\end{eqnarray}
hold at least for quark masses up to that of the strange quark.
Our results for the strange-particle spectrum are encouragingly
close to the experimental values, and the dependence of both the
pseudoscalar and vector decay constants on $m_s$ is consistent
with experiment.  The only serious problem in the meson sector
shows up in the actual values of the decay constants.  For the
pseudoscalar case, the sign of the discrepancy is consistent with
the expectation that the wavefunction at the origin is reduced in
the quenched approximation, but whether the magnitude is entirely
attributable to this effect or is due in part to our use of the
perturbative value for the axial current renormalisation is not
known.  Our results for the vector decay constants are more
difficult to interpret, as the discrepancy is in the opposite
direction, which may signal the presence of other effects.  It is
clearly important to calculate the current renormalisations
non-perturbatively at $\beta=6.2$ for at least two values of the
quark mass.

We have not explored $SU(3)$-flavour-symmetry breaking in the
baryon sector, but only present results for degenerate quarks.
Compared with the chiral extrapolations for the other hadrons, a
linear chiral extrapolation for the nucleon is not well supported
by our data, and leads to a nucleon mass which is lower than is
typical of quenched simulations to date, although in better
agreement with experiment! On the other hand, our results for the
$\Delta$ and $\Omega$ are less problematic.  They are also in good
agreement with experiment, and the result for the $\Omega$, along
with our results for mesons, encourages us to believe that we have
good control of the strange-quark physics that can be obtained
from 2-point functions.


\paragraph{Acknowledgements}

This research is supported by the UK Science and Engineering Research
Council under grants GR/G~32779, GR/H~49191, GR/H~53624 and
GR/H~01069, by the University of Edinburgh and by Meiko Limited.  CTS
and ADS thank SERC for financial support.  We are grateful
to Edinburgh University Computing Service and, in particular, to Mike
Brown for his tireless efforts in maintaining service on the Meiko
i860 Computing Surface.




\end{document}